\def\preprint{0}                
\def\preprint{1}                
\def\comment#1{}
\if\preprint1
        \documentclass[usenatbib]{mn2e}
        \usepackage{times}
        \usepackage{graphicx}
        \usepackage{calc}

\else
        \documentstyle[astrop-bib,referee,times]{mn}
        \newcommand{\includegraphics}[1]{}
\fi

\def\oversim#1#2{\lower0.5pt\vbox{\baselineskip0pt \lineskip-0.5pt
     \ialign{$\mathsurround0pt #1\hfil##\hfil$\crcr#2\crcr\sim\crcr}}}
\def\lsim{\mathrel{\mathpalette\oversim<}}    

\title[Mass-losing AGB stars in the LMC]
{A Spitzer mid-infrared spectral survey of mass-losing carbon stars in the 
Large Magellanic Cloud}
\author[A.A. Zijlstra et al.]{
               Albert~A.~Zijlstra$^1$\thanks{E-mail: 
                   \tt a.zijlstra@manchester.ac.uk},
       Mikako Matsuura$^{1,2}$,
       Peter R. Wood$^3$,  G.C. Sloan$^4$, Eric Lagadec$^1$, 
         \newauthor 
       Jacco Th. van Loon$^5$,
       M.A.T. Groenewegen$^6$,  
       M.W. Feast$^7$, J.W. Menzies$^8$, P.A. Whitelock$^8$,
        \newauthor
       J.A.D.L. Blommaert$^6$, M.-RL. Cioni$^9$, H.J. Habing$^{10}$, 
       S. Hony$^6$, C. Loup$^{11}$, 
       L.B.F.M. Waters$^{12}$\\
        $^1$University of Manchester, School of Physics \&\ Astronomy, 
          P.O. Box 88, Manchester M60 1QD, UK\\
        $^2$Department of Pure and Applied Physics, The Queen's University of 
        Belfast, Belfast BT7 1NN, UK\\
        $^3$Research School of Astronomy and Astrophysics, 
           Australian National University,
            Cotter Road, Weston Creek, ACT 2611, Australia\\
        $^4$Department of Astronomy, Cornell University, 108 Space Sciences 
            Building, Ithaca NY 14853-6801, USA\\
        $^5$Astrophysics Group, School of Physical \&\ Geographical Sciences, 
           Keele University, Staffordshire ST5 5BG, UK\\
        $^6$Instituut voor Sterrenkunde, K.U. Leuven,
          Celestijnenlaan 200B, B-3001 Leuven, Belgium\\ 
        $^7$ Astronomy Department, University of Cape Town, 7701 Rondebosch, 
            South Africa\\
        $^8$South African Astronomical Observatory, PO Box 9, 7935
          Observatory,  South Africa\\
        $^9$Institute for Astronomy, University of Edinburgh, Royal 
           Observatory, Blackford Hill, Edinburgh EH9 3HJ, UK\\
        $^{10}$Sterrewacht Leiden, Niels Bohrweg 2, 2333 RA Leiden, 
            The Netherlands\\
        $^{11}$Institut d'Astrophysique de Paris, CNRS, 98bis Boulevard Arago,
               75014 Paris, France\\
        $^{12}$Astronomical Institute, University of Amsterdam, Kruislaan 403, 
            1098 SJ Amsterdam, The Netherlands\\
}
\pubyear{2005}

\begin{document}

\maketitle
\begin{abstract}
  We present a {\it Spitzer Space Telescope} spectroscopic survey of
  mass-losing carbon stars (and one oxygen-rich star) in the Large Magellanic
  Cloud.  The stars represent the superwind phase on the Asymptotic Giant
  Branch, which forms a major source of dust for the interstellar medium in
  galaxies.  The spectra cover the wavelength range 5--38\,$\mu$m. They show
  varying combinations of dust continuum, dust emission features (SiC, MgS)
  and molecular absorption bands (C$_2$H$_2$, HCN).  A set of four narrow
  bands, dubbed the Manchester system, is used to define the infrared
  continuum for dusty carbon stars.  The relations between the continuum
  colours and the strength of the dust and molecular features are studied, and
  are compared to Galactic stars of similar colours. The circumstellar
  7-$\mu$m C$_2$H$_2$ band is found to be stronger at lower metallicity, from
  a comparison of stars in the Galaxy, the LMC and the SMC.  This is explained
  by dredge-up of carbon, causing higher C/O ratios at low metallicity (less
  O).  A possible 10-$\mu$m absorption feature seen in our spectra may be due
  to C$_3$. This band has also been identified with interstellar silicate or
  silicon-nitrite dust. We investigate the strength and central wavelength of
  the SiC and MgS dust bands as function of colour and metallicity. The
  line-to-continuum ratio of these bands shows some indication of being lower
  at low metallicity.  The MgS band is only seen at dust temperatures below
  600\,K.  We discuss the selection of carbon versus oxygen-rich AGB stars
  using the J$-$K vs. K$-$A colours, and show that these colours are
  relatively insensitive to chemical type.  Metal-poor carbon stars form
  amorphous carbon dust from self-produced carbon. This type of dust forms
  more readily in the presence of a higher C/O ratio. Low metallicity carbon
  dust may contain a smaller fraction of SiC and MgS constituents, which do
  depend on metallicity. The formation efficiency of oxygen-rich dust depends
  more strongly on metallicity. We suggest that in lower-metallicity
  environments, the dust input into the Interstellar Medium by AGB stars is
  efficient but may be strongly biassed towards carbonaceous dust, as compared
  to the Galaxy.
\end{abstract}

\begin{keywords}
 Galaxies: Magellanic Clouds;
 stars: stars: AGB and post-AGB
 -- stars: carbon
 -- stars: mass loss
 -- stars: infrared
 dust
\end{keywords}

\section{Introduction}

Low and intermediate-mass stars ($\rm M\sim 1$--$8\,\rm M_\odot$, hereinafter
LIMS) make up $>$90\% of all the stars which have died in the Universe up to
the present time.  During their late evolution, after entering the high
luminosity asymptotic giant branch (AGB) phase, the stars eject their
hydrogen-rich outer layers during a phase of catastrophic mass loss: the
so-called superwind phase.  After this final burst of activity, the star
remains as a hot, compact white dwarf of mass 0.6--1.4 M$_{\odot}$. The
expanding ejecta surrounding the star become ionized and form a planetary
nebula, before dispersing into the interstellar medium.

During the superwind, the star tends to be self-obscured by dust which forms
in the ejecta. Dredge-up of primary carbon (produced by triple-$\alpha$
burning) may turn the star into a carbon star. At the low effective
temperature of AGB stars, the ejecta are largely molecular.  The first and
most stable molecule to form is CO: this locks up the least abundant of the C
and O atoms. Other molecules and dust form from the remaining atoms. The
chemistry changes dramatically as the C/O ratio passes unity, and this affects
both the gas-phase species \citep{Millar2000, Willacy1997} and the dust
composition \citep{Treffers1974}.  In carbon stars, the C/O ratio (by number)
is more than unity, and carbon-rich molecules (e.g. C$_2$H$_2$, HCN) and
carbonaceous dust \citep{Hony2002a} result, whilst in oxygen stars (C/O$<1$),
metal oxides and silicate dust \citep{Cami2002} form.

The mass-loss process is important for three reasons. First, it determines the
mass distribution of stellar remnants, including e.g. the lower mass limit of
type-II supernovae progenitors \citep{Zijlstra2004a}.  Second, the physics and
interplay of stellar pulsation, shock waves, dust formation and radiation
pressure which drives the mass loss are still little understood. Third,
stellar mass loss drives galactic evolution through replenishment and
enrichment of the Interstellar Medium (ISM). On this last point, mass-loss
from LIMS contributes roughly half the total gas recycled by all stars
\citep{Maeder1992}, creates an amount of carbon roughly equal to that produced
by supernovae and Wolf-Rayet stars \citep{Dray2003, Gavilan2005} and is the
main source of carbonaceous interstellar dust \citep{Edmunds2001, Dwek1998}.
LIMS are the only confirmed source of the primary nitrogen required to explain
observed abundances in objects in the early universe, are the main source of
$^{13}$C \citep{Hajduk2005} and heavy s-process elements \citep{Goriely2000,
  Wanajo2006}, and are the major stellar source of lithium \citep{Romano2001}.

\begin{figure*}
\includegraphics[width=16cm]{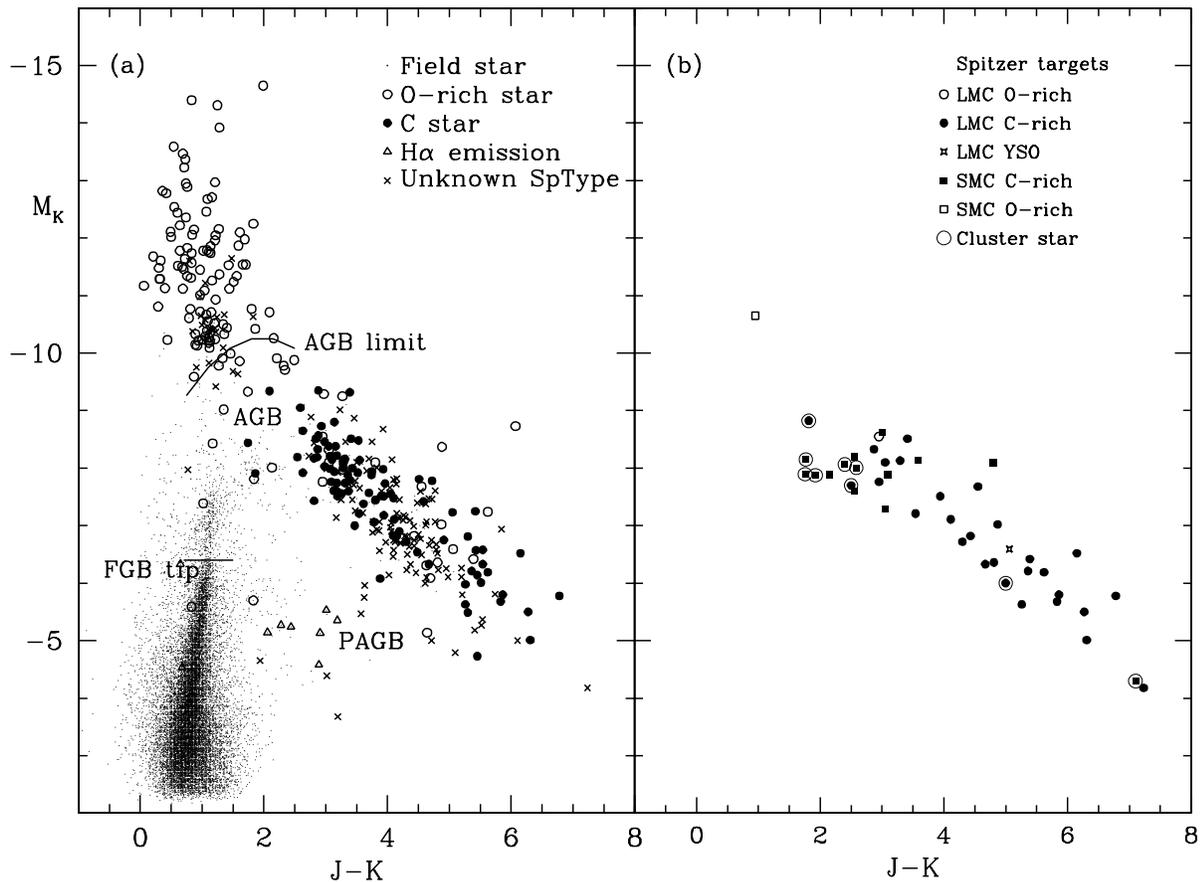}
\caption{\label{colours.eps}  (a): The M$_{K}$, $J$-$K$ diagram for a sample 
  of known mid-IR sources (MSX and IRAS) in the LMC (large points), together
  with field stars (small symbols) from an area $\sim 60$ times smaller than
  the area from which the mid-IR sources were selected.  AGB stars are
  confined approximately to the region below the line marked ``AGB limit''.
  Stars above this limit are foreground stars and supergiants: the supergiant
  stars have masses $M > 8$M$_{\odot}$. Spectral types were those available
  before the Spitzer observations were made (b): similar to (a), except that
  te objects shown are the point sources actually observed, and the spectral
  types come from recent ground-based and Spitzer spectra.  In both panels,
  distance modulii of 18.5 and 18.9 have been assumed for the LMC and SMC,
  respectively.  J and K photometry is from various sources existing prior to
  the Spitzer observations.
 }
\end{figure*}

The AGB and post-AGB evolution is dominated by the superwind mass loss.
Theoretical models \citep[e.g.][]{Sandin2003} cannot yet predict mass-loss
rates from stellar parameters: instead observational knowledge is required of
the mass-loss rates as a function of mass, luminosity and metallicity.
However, the observational relations are themselves not well calibrated: local
Galactic stars have poorly known distances and unknown progenitor masses,
while extra-galactic stars are in general too faint to detect the mass-loss
tracers. Important work has been done on Bulge stars \citep{Ortiz2002}.  The
main tracer for the superwind is the dust, but dust emits in the thermal
infrared. Surveys by the Infrared Astronomical Satellite (IRAS) uncovered a
number of luminous self-obscured AGB stars in the Magellanic Clouds
\citep{Reid1991, Zijlstra1996}. A larger number of these stars was discovered
more recently with the Midcourse Space Experiment (MSX) satellite
\citep{Egan2001}.  The Infrared Space Observatory (ISO) made it possible to
measure mass-loss rates, but only for the most massive stars and the highest
mass-loss rates \citep[e.g.][]{vanLoon1999a}.

The {\it Spitzer Space Telescope} \citep{Werner2004} for the first time
provides the sensitivity necessary to obtain mass-loss rates for the whole
range of AGB masses and luminosities in the Magellanic Clouds.  These
stars have well-known distances, meaning that absolute luminosities can be
easily obtained. The general abundance of field stars may be estimated from
the well-studied age-metallicity relations, and abundances have been derived
explicitly for many star clusters. We have thus undertaken a survey of stars
all along the AGB branch in the Large Magellanic Cloud and the Small
Magellanic Cloud (hereinafter LMC and SMC respectively) with {\it Spitzer}. In
this paper we present the LMC stars observed in our survey. Data for the SMC
stars and the analysis of mass-loss rates will appear in subsequent papers.

\section{Target selection}

\begin{figure}
\includegraphics[width=8cm,clip=true]{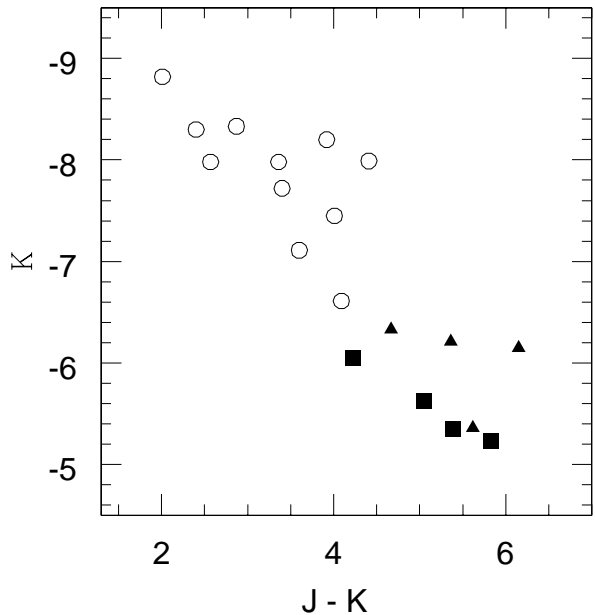}
\caption{\label{k_jk.ps}  The $M_{\rm K}$, $\rm J-K$ diagram for the sources
  in this paper. J and K values are taken from Tables \ref{targets.dat} and
 \ref{periods.dat}, and from \citet{vanLoon2005a}. Objects with lower limits
 only for the J$-$K colour are not included. Symbols indicate a source
 classification based on data described in this paper: open symbols: stars
 without a MgS feature, filled triangles: a weak SiC features and MgS, filled
 squares: strong SiC and MgS }
\end{figure}

The ($M_{\rm K}$,\,$\rm J-K$) diagram in Fig.~\ref{colours.eps}(a) shows most
of the AGB stars which are known mid-IR point sources in the LMC.  Also shown
is a representative sample of the much more numerous field stars without thick
dust shells.  The most numerous stars in any stellar population are low-mass
stars.  In Fig.~\ref{colours.eps}(a), these stars evolve up the AGB in $M_{\rm
  K}$ at $\rm J-K < 2.5$ and $M_{\rm K}\!>\!-9$, before they develop high
mass-loss rates and thick dust shells and evolve to $\rm J-K \sim 7$ and
$M_{\rm K} \sim -5$.  Fig.~\ref{colours.eps}(a) also includes massive AGB
stars ($M\!>\!3\,\rm M_{\odot}$, $-10.5\!<\!M_{\rm K}\!<\!-9$), deriving from
intermediate-mass progenitor stars.

Stars observed in the current survey are shown in Fig.  \ref{colours.eps}(b).
We selected targets in both the SMC and the LMC from the {\it Midcourse Space
  Experiment} (MSX) catalog \citep{Egan2001}, the IRAS point source catalog,
the ISO catalog of variable AGB stars in the SMC \citep{Cioni2003},
long-period variables in the Magellanic Clouds \citep[e.g.][]{Wood1983,
  Feast1989}, and the intermediate-age clusters NGC419 and NGC1978, both of
which have a known mid-infrared source.  Targets were selected to span the
long sequence of increasing mass loss in LIMS which stretches from $(\rm
J\!-\!K,M_{\rm K})\sim(2,-9)$ to (7,$-4$) in Fig.  \ref{colours.eps}(a).  (The
bright O-rich object near $(\rm J\!-\!K,M_{\rm K})\sim(1,-10.7)$ in Fig.
\ref{colours.eps}(b) was not a target but is an object which was accidently
observed when {\it Spitzer} selected it in the peak-up field). It turns out
that only two of the LMC sources were oxygen rich, one of which was found to
be a massive young stellar object \citep{vanLoon2005b}.  Here we present the
carbon-rich sources in the LMC. The SMC sample will be presented elsewhere.

 Table \ref{targets.dat} lists the 29 stars included here.  The objects have
been included in various catalogues over time, accumulating different
names. We adopt the oldest available name for each object, but
Table~\ref{targets.dat} gives the cross references to names used by
\citet{Reid1990}, \citet{Egan2001} and in the MSX catalogue. The adopted
positions are from 2MASS counterparts and differ slightly
from the MSX positions.

\begin{table*}
\caption[]{\label{targets.dat} Observed LMC targets:
names, adopted coordinates,  and  photometry.
The MSX photometry is for band A (8.2 micron) and refers to version 2.3
of the catalogue. The zeropoint for the MSX band A
is taken as 58.46\,Jy \citep{Egan2001}.
JHK is taken from 2MASS, except for MSX LMC 494
and NGC1978 MIR1 where the values are from \citep{Ita2004b} and 
\citep{vanLoon2005a} respectively.  $M_{\rm bol}$ (Section \ref{mbol}) 
assumes  a distance modulus of 18.5.
  }
\begin{flushleft}
\begin{tabular}{lllllrrrrlllllll}
\hline
Adopted name &  MSX source & other name & RA & Dec  
& \multicolumn{1}{c}{J} & \multicolumn{1}{c}{H} & \multicolumn{1}{c}{Ks} &  \multicolumn{1}{c}{A}
 & \multicolumn{1}{c}{$M_{\rm bol}$}    \\
  & & &  \multicolumn{2}{c}{(J2000)} & \multicolumn{1}{c}{mag} & \multicolumn{1}{c}{mag} &
  \multicolumn{1}{c}{mag} & \multicolumn{1}{c}{mJy}  \\
\hline
{\it carbon rich} \\
MSX LMC 219       & 051119.5$-$684227 &           & 05 11 19.52 & $-$68 42 27.9
& $>$15.61 & $>$14.67 & 13.14 & 204 & $-4.76$ \\
IRAS 05132$-$6941 & 051250.8$-$693749 & MSX LMC 223 &  05 12 51.07 & $-$69 37 50.3 
& $>$17.81 & $>$15.68 & 13.36 & 259 & $-5.01$  \\
MSX LMC 349       & 051726.9$-$685458 &           & 05 17 26.94 & $-$68 54 58.2 
& $>$17.44 & $>$16.88 & 14.82 & 193 & $-4.97$\\
MSX LMC 341       & 052100.5$-$692054 &           & 05 21 00.37 & $-$69 20 55.3 
& $>$17.49 & 15.9\rlap{:} & 13.151 & 210  & $-4.74$  \\
MSX LMC 663       & 052244.5$-$693826 &           & 05 22 44.00 & $-$69 38 28.1 
& 12.60 & 11.33 & 10.20 & 75  & $-5.40$  \\
MSX LMC 494       & 052309.6$-$691744 &           & 05 23 09.11 & $-$69 17 49.1 
& 13.08  &  11.71 & 10.51  & 94 & $-4.98$ \\
MSX LMC 441       & 052438.7$-$702357 &           & 05 24 38.62 & $-$70 23 57.1 
& $>$17.16 & $>$16.44 & 14.48 & 200 & $-4.92$  \\
MSX LMC 443       & 052505.9$-$701011 &           & 05 25 05.69 & $-$70 10 10.6 
& $>$14.54 & 13.72 & 11.54 & 150 & $-4.68$ \\
MSX LMC 601       & 052650.9$-$693136 &           & 05 26 50.83 & $-$69 31 36.9 
& 13.88 & 11.99 & 10.52  & 172 & $-5.14$ \\
NGC 1978 IR4      &       &       WBT 815       & 05 28 44.50 & $-$66 14 04.0 
& 11.49\rlap{:} & 10.34\rlap{:} & 9.68 & 70\rlap{:}  & $-5.35$ \\
NGC 1978 MIR1     &                 &           & 05 28 47.20 & $-$66 14 13.6 
& 16.67 &    & 12.45  & 124 & $-4.75$ \\
NGC 1978 IR1 &            &       WBT 1268      &  05 28 40.17 & $-$66 13 54.2 
& $>$13.78 & $>$12.78 & 11.73 & 77 &  $-5.03$ \\
IRAS 05278$-$6942 & 052724.3$-$693944 & MSX LMC 635 & 05 27 24.12 & $-$69 39 45.0 
& $>$17.53 & $>$15.59 & 12.35 & 683 & $-6.57$ \\
MSX LMC 754       & 052811.4$-$703359 &           & 05 28 11.48 & $-$70 33 58.7 
& 15.98\rlap{:} & 13.70 & 11.89 & 93\rlap{:} & $-4.91$\\
MSX LMC 679       & 052848.5$-$694801 &            & 05 28 48.62 & $-$69 48 01.3 
& $>$15.84 & 14.57 & 12.48 & 121 & $-4.93$  \\
MSX LMC 743       & 053454.1$-$702925 &            & 05 34 53.74 & $-$70 29 24.8 
& $>$14.13 & $>$13.22 & 13.27 & 165 & $-5.27$ \\
MSX LMC 749       & 053527.1$-$695229 &            & 05 35 26.86 & $-$69 52 27.9 
& 15.06 & 12.93 & 11.05 & 142 &  $-5.56$ \\
MSX LMC 967       & 053637.2$-$694725 &            & 05 36 36.71 & $-$69 47 22.6 
& 13.48 & 12.00 & 10.67 & 87\rlap{:} &   $-5.45$ \\
IRAS 04557$-$6753 &                 & MSX LMC 1238 & 04 55 38.98 & $-$67 49 10.7 
& $>$16.18  & 14.49 &  12.397& 157 & $-5.05$ \\
IRAS 05009$-$6616 &                 & MSX LMC 1278 & 05 01 04.43 & $-$66 12 40.4 
& $>$15.64 & 14.57 & 12.40 & 240 & $-5.57$ \\
IRAS 05291$-$6700 &       &     GRRV 38          &  05 29 07.60 & $-$66 58 15.0 
& 12.45 & 10.91 & 9.91 & 81\rlap{:} & $-4.91$ \\
IRAS 05295$-$7121 &                 & MSX LMC 692  & 05 28 46.62 & $-$71 19 12.5 
& 16.80\rlap{:} &  14.32 &  12.18 & 204 & $-5.24$ \\
TRM 88            &                 & MSX LMC 310  & 05 20 19.38 & $-$66 35 47.8 
& 14.76 & 12.81 & 11.10 & 169 & $-5.15$ \\
IRAS 05112$-$6755 &                 & MSX LMC 44, TRM4 & 05 11 10.47 & $-$67 52 10.5 
& 16.41\rlap{:} & 14.07 & 11.69 & 451 & $-5.59$ \\
IRAS 05190$-$6748 &                 & MSX LMC 307, TRM20 &  05 18 56.26 &
$-$67 45 04.4 
& $>$18.21 & $>$15.91 & 13.10 & 209 & $-5.61$ \\
IRAS 05113$-$6739 &                 & MSX LMC 47, TRM24 & 05 11 13.89 & $-$67 36 16.1 
& $>$17.72 & 14.77 & 12.49 & 233 & $-5.18$ \\
TRM 72             &                 & MSX LMC 29
                      & 05 11 38.65 & $-$66 51 09.8 
& 14.34 & 12.36 & 10.74 & 122 & $-5.43$ \\
IRAS 05360$-$6648 &                 & MSX LMC 872, TRM77\quad & 05 36 01.24 & $-$66 46 39.7 
& $>$18.04 & 16.07\rlap{:} & 13.28 & 192 & $-4.73$ \\
\hline 
{\it oxygen rich} \\
IRAS 05003$-$6712 &                 & MSX LMC 1280 & 05 00 19.00 & $-$67 07 58.0 
 & 12.04 & 10.46 & 9.32 & 205 & $-5.42$ \\
\hline \\
\end{tabular}
\end{flushleft}
\end{table*}

The JHKs photometry in Table \ref{targets.dat} is taken from the 2MASS All-sky
catalog.  It is based on a single epoch only and does not sample the
variability.  Where further data is available, we list JHKL photometry
averaged over the cycle in Table \ref{periods.dat}.  Also listed are pulsation
periods, taken from \citet{Whitelock2003}, \citet{Wood1998},
\citet{Ishida2000}, \citet{Ita2004b} and \citet{Groenewegen2004}. Where
available, periods from \citet{Whitelock2003} are listed in preference because
of the larger number of observations used.  Photometry is taken from the first
three sources, with values from Wood listed in preference, where available.
Especially the L-band filters differ substantially between the sources; the
photometry is in the original system.  For MSX LMC 967, the photometry is an
average of two observations only: 2MASS and a single measurement from
\citet{Ita2004b}.

Fig. \ref{k_jk.ps} shows the $\rm M_K$ versus $\rm J - K$ colour-magnitude
diagram, using photometry listed above supplemented with data from
\citet{vanLoon2006a}.  The stars follow the line of increasing mass loss
defined in Fig.  \ref{colours.eps}(a).  Neither figure  includes the
stars which are faintest at K, as these show only uper limits for J.

\begin{table}
\caption[]{\label{periods.dat} Phase-averaged photometry and
periods. Periods are taken from \citet{Whitelock2003},
\citet{Wood1998}, \citet{Ishida2000},  and \citet{Groenewegen2004}.  }
\begin{flushleft}
\begin{tabular}{llllll}
\hline
Adopted name 
& J & H & K &  L & P   \\
  &  mag & mag & mag & mag & days \\
\hline
MSX LMC 663       & & & & & 455: \\
MSX LMC 494       & & & & & 458 \\
MSX LMC 601       & & & & & 546 \\
NGC 1978 IR1      & 14.26 & 12.01 & 10.51 & -- & 491 \\
MSX LMC 749       & & & & & 329, 589 \\
MSX LMC 967       & (14.2) & (12.5) & (10.8) & -- & 572 \\
IRAS 04557$-$6753 &        & 13.97 & 11.78 & 9.11 & 765 \\
IRAS 05009$-$6616 &        & 13.06 & 11.29 & 8.94 & 658 \\
GRRV 38           &  13.04 & 11.33 & 10.17 & 8.86 & 483 \\
IRAS 05295$-$7121 &        & 12.96 & 10.99 & 8.83 & 682 \\
TRM 88            &  14.22 & 12.10 & 10.30 & 8.26 & 544 \\
IRAS 05112$-$6755 &  17.65 & 15.11 & 12.29 & 8.89 & 830 \\
IRAS 05190$-$6748 &        & 16.11 & 12.72 & 8.71 & 939 \\
IRAS 05113$-$6739 &  16.84 & 14.63 & 12.17 & 8.93 & 700 \\
TRM 72            &  15.50 & 13.33 & 11.39 & 8.92 & 631 \\
IRAS 05360$-$6648 &        & 15.50 & 12.87 & 9.24 & 538 \\
\hline
IRAS 05003$-$6712 &  12.90 & 11.26 & 9.95  & 8.49 & 883 \\
\hline \\
\end{tabular}
\end{flushleft}
\end{table}

\subsection{Comments on individual stars}

Three stars in Table \ref{targets.dat} are located in the cluster NGC 1978, as
indicated by their names. The cluster was studied by \citet{Tanabe1997}; it
has an age of $2.5\, 10^9\,$yr, a turn-off mass of about 1.5$\,\rm M_\odot$
and a metallicity [Fe/H]$\,=-0.66\pm0.22$ \citep{vanLoon2005a}. One star,
MIR-1, is a known obscured LPV, one other is an MSX-detected star, and the
third \citep[IR4,][]{vanLoon2005a,Will1995} is an AGB star with no listed MSX
point-source counterpart.  This last star is seen on MSX images, from which
\citet{vanLoon2005a} estimate a flux density of 70 mJy in MSX band A (6.8--10.8
micron).

IRAS 04557$-$6753 is also located in a (small) cluster, KMHK 285
\citep{Bica1999, vanLoon2005a}, but this cluster is little studied. The age is
estimated as $>10^9\,\rm yr$ and the metallicity is not known.

MSX LMC 494 is identified with a large amplitude variable observed with the
Optical Gravitational Lensing Experiment, OGLE~052309.18$-$691747.0.

LMC MSX 663 is also a long-period OGLE variable but its OGLE amplitude is very
small \citep{Groenewegen2004}.  It is an optically bright AGB star previously
classified as an S star \citep{Cioni2001}. However, optical spectra show that
it is a symbiotic carbon star with strong emission lines (Wood et al, in
preparation).

IRAS 05291$-$6700 was identified with a nearby variable star, GRRV~38
\citep{Glass1985}, but the coordinates of the IRAS source differ by 30 arcsec
from this star.  The IRAS point source is likely unrelated.  There is confusion
with an extended mid-infrared source approximately 6 arcminutes east.  MSX
only has the stellar source, but nothing at the position of the IRAS source.
The identification of the star \citep{Zijlstra1996} most likely occurred
serendipitously. GRRV~38 is a carbon star \citep{vanLoon1999b}; the infrared
colours indicate little or no obscuration. Below we will use the name GRRV 38
for this object.

For TRM~88, \citet{Wood1998} and \citet{Whitelock2003} report main periods of
565 and 544 days respectively.  The light curve is affected by long-term
secular changes  \citep{Whitelock2003} due to variable circumstellar
extinction \citep{Feast2003}.

IRAS\,05009$-$6616 and TRM\,72 also show evidence for long-term variations in
the light curves \citep{Whitelock2003}.

TRM\,72 should not be confused with the well-studied CH star WORC~106
\citep{Feast1992} which is located 30 arcsec away.

\section{Observations}

We observed our targets with the Infrared Spectrograph (IRS)
\citep{Houck2004} on the {\it Spitzer Space Telescope} using the
low-resolution modules Short-Low (SL) and Long-Low (LL), which
together cover wavelengths from 5 to 38\,$\mu$m.  The slits in SL
are roughly perpendicular to LL, and each module has two
separate apertures, with the longer wavelengths observed in
first order and the lower in second order.  In order
of increasing wavelength, the four spectral segments are
known as SL2, SL1, LL2, and LL1.  In addition, a ``bonus''
order covers the overlap between the first and second orders
in each module; this order is a small piece of the first-order
spectrum obtained when the source is in the second-order
aperture.  When the source is in one aperture, the other
aperture is exposed to a section of nearby sky, providing
a background measurement, e.g. in SL2 when the source is in
the SL1 aperture.  Each source was observed in two nod
positions in each aperture.  The resolution varies between
64 and 128, depending on the order and wavelength.  The slit
dimensions are approximately 3.5$\arcsec$ $\times$ 57$\arcsec$
for SL and 10.5$\arcsec$ $\times$ 168$\arcsec$ for LL (width
$\times$ length).

Different exposure times were used depending on expected target flux.
For  SL, 12 to 120-sec exposure times were used per segment. For LL1,
exposure times ranged from 60 to 960 sec, and for LL2 from 180 to 960 sec.
Observations were taken in standard staring mode.

Flat-fielded images were generated by the {\it Spitzer} Science Center (SSC).
We replaced bad pixels with values estimated from neighbouring pixels (using
the {\it imclean.pro} package), and removed the background emission by
differencing images aperture-by-aperture in SL (e.g. SL1$-$SL2 and vice
versa), and nod-by-nod in LL.  Spectra were extracted from the images using
the software provided with the Spitzer IRS Custom Extractor (SPICE) available
from the SSC.  They were calibrated making a full spectral correction
using HR 6348 (K0 III) as a standard star in SL and HR 6348, HD 166780 (K4
III), and HD 173511 (K5 III) in LL.
Individual images were extracted separately and then co-added.  Nods were
combined, then spectral segments were combined by making scalar multiplicative
corrections for pointing-induced throughput errors (normalizing to the best
centered segment).  The accuracy of the overall spectral shape is limited to a
few percent. The matching across the different sections (e.g. LL2 to LL1) may
leave some residuals.

The standard wavelength calibration is accurate to 0.06 micron in SL and 0.15
micron in LL.  Flux calibration is accurate to $\lsim$ 5 per cent. The
response curve is known to a few per cent, but residuals at this level may
remain. The largest residual calibration errors are due to pointing
uncertainties.  The LL array is affected by fringing which in most cases
has been corrected. More details on calibration issues can be found
in \citet{Sloan2003}.

\begin{figure*}
\begin{center}
\includegraphics[height=22cm]{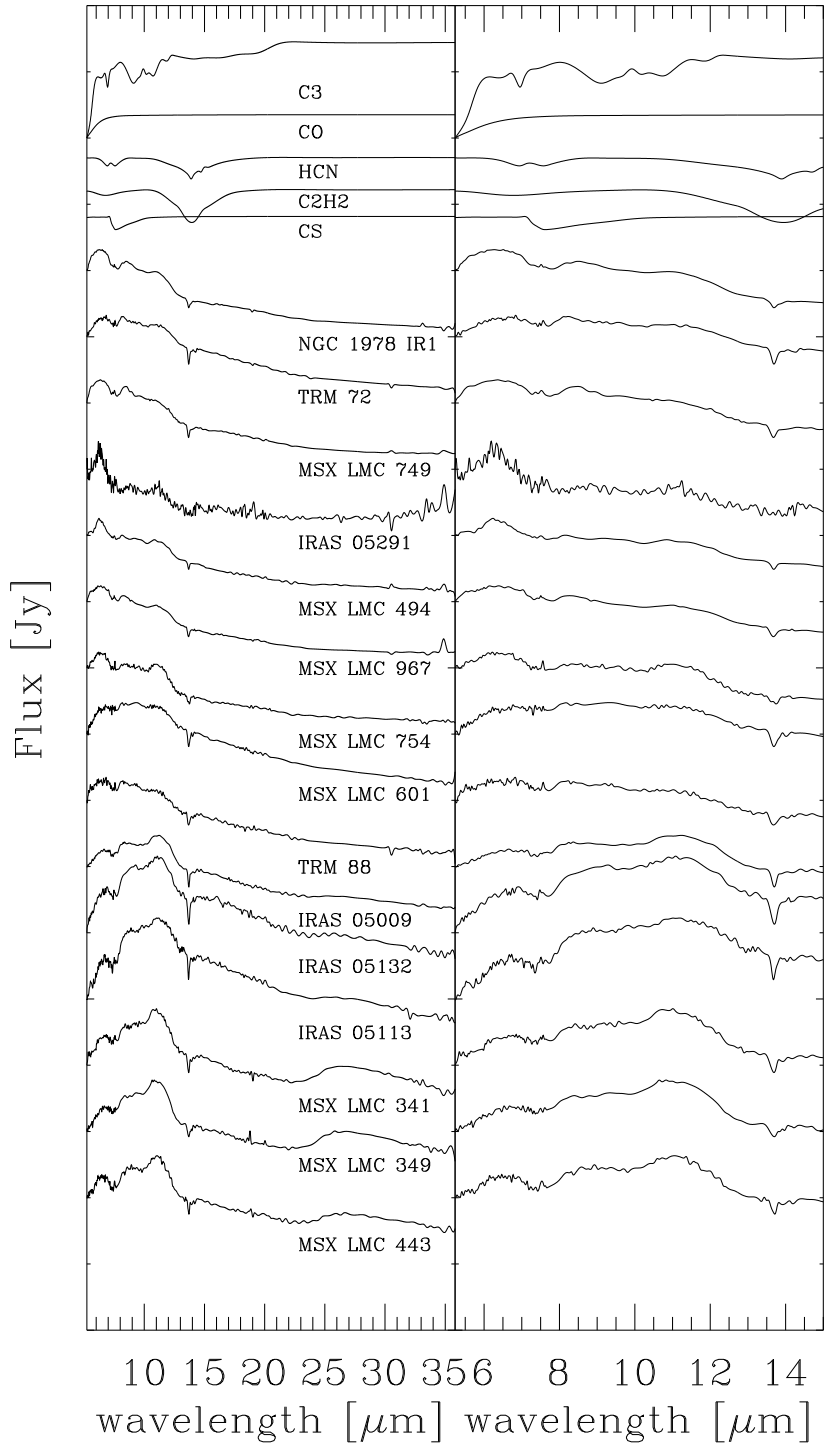}
\caption{\label{spectra1} Spectra of the 15 bluer carbon-rich AGB stars 
observed in the LMC. The top five lines show the model molecular spectra
from \citet{Jorgensen2000} }
\end{center}
\end{figure*}

\begin{figure*}
\includegraphics[height=22cm]{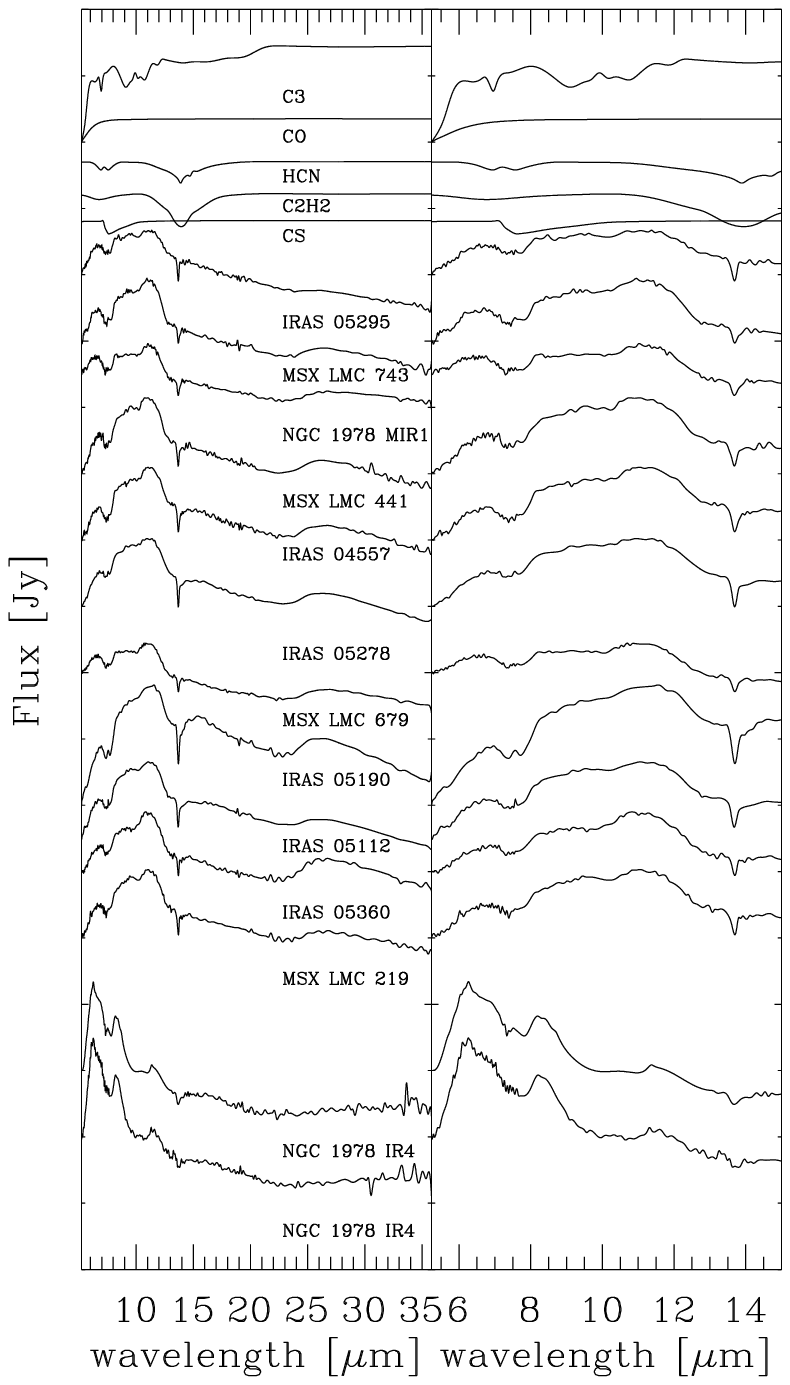}
\caption{\label{spectra2} Spectra of 11 redder carbon-rich AGB stars 
observed in the LMC, and  spectra of the blue ('naked') carbon star
 NGC\,1978\,IR4 which was observed twice.  The top five lines show the 
model molecular spectra from \citet{Jorgensen2000} }
\end{figure*}

The spectra of the carbon stars are shown in Fig.~\ref{spectra1} and
~\ref{spectra2}. They are ordered by dust temperature, from blue spectra to
red. The spectra show numerous absorption and emission bands of molecules and
dust resonances, superposed on the dust continuum.  We will first discuss
these bands, separated into dust and molecular features. The following section
will discuss the continuum spectral energy distribution.

Some objects show residual interstellar lines, due to incomplete cancellation
of the background emission. These lines are found at 33.5\,$\mu$m ([S\,{\sc
  III}]), 34.8\,$\mu$m ([Si\,{\sc II}]), and 36.0\,$\mu$m ([Ne\,{\sc  III}]).
An apparent feature at 30.5\,$\mu$m seen in a few spectra is an artifact,
due to a hot pixel.

\begin{figure}
\includegraphics[width=8cm]{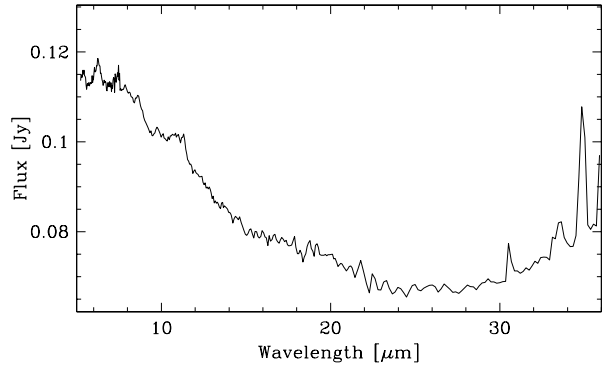} 
\caption{\label{spectra3} Spectrum of the symbiotic C star MSX LMC 663 }
\end{figure}

The object MSX LMC 663 exhibits an IRS spectrum which is very different from
the others, as shown in Fig. \ref{spectra3}. It is affected by interstellar
emission in the background measurement, causing the offset measurements to
disagree; however, the rise beyond 25\,$\mu$m appears to be intrinsic to the
star and not due to background emission.  Shortward, the spectrum shows little
evidence for dust but the absorption bands are different in shape from
those of the other carbon stars.

IRAS\,05003$-$6712 is the only O-rich AGB star in our sample \citep[one other
object was classified as a massive young stellar object, and is published in
][]{vanLoon2005b}. It has an OH maser \citep{Marshall2004} but only the
blue-shifted component is detected. The {\it Spitzer} spectrum is dominated by
the silicate emission bands at 10 and 18\,$\mu$m, with perhaps a hint of
CO$_2$ absorption at 13.5\,$\mu$m and SiO absorption at 8.6\,$\mu$m.
\citep[SiO bands tend to be weak in LMC Miras: ][]{Matsuura2005a}. The
10-$\mu$m feature is typical for a lower mass-loss rate: it is comparable to
that of the LMC AGB star IRAS\,04544$-$6849 \citep{Dijkstra2005} with a
mass-loss rate of $\dot M = 7 \times 10^{-6}\,\rm M_\odot\,yr^{-1}$
\citep{vanLoon1999a}. .

\begin{figure}
\includegraphics[width=8cm]{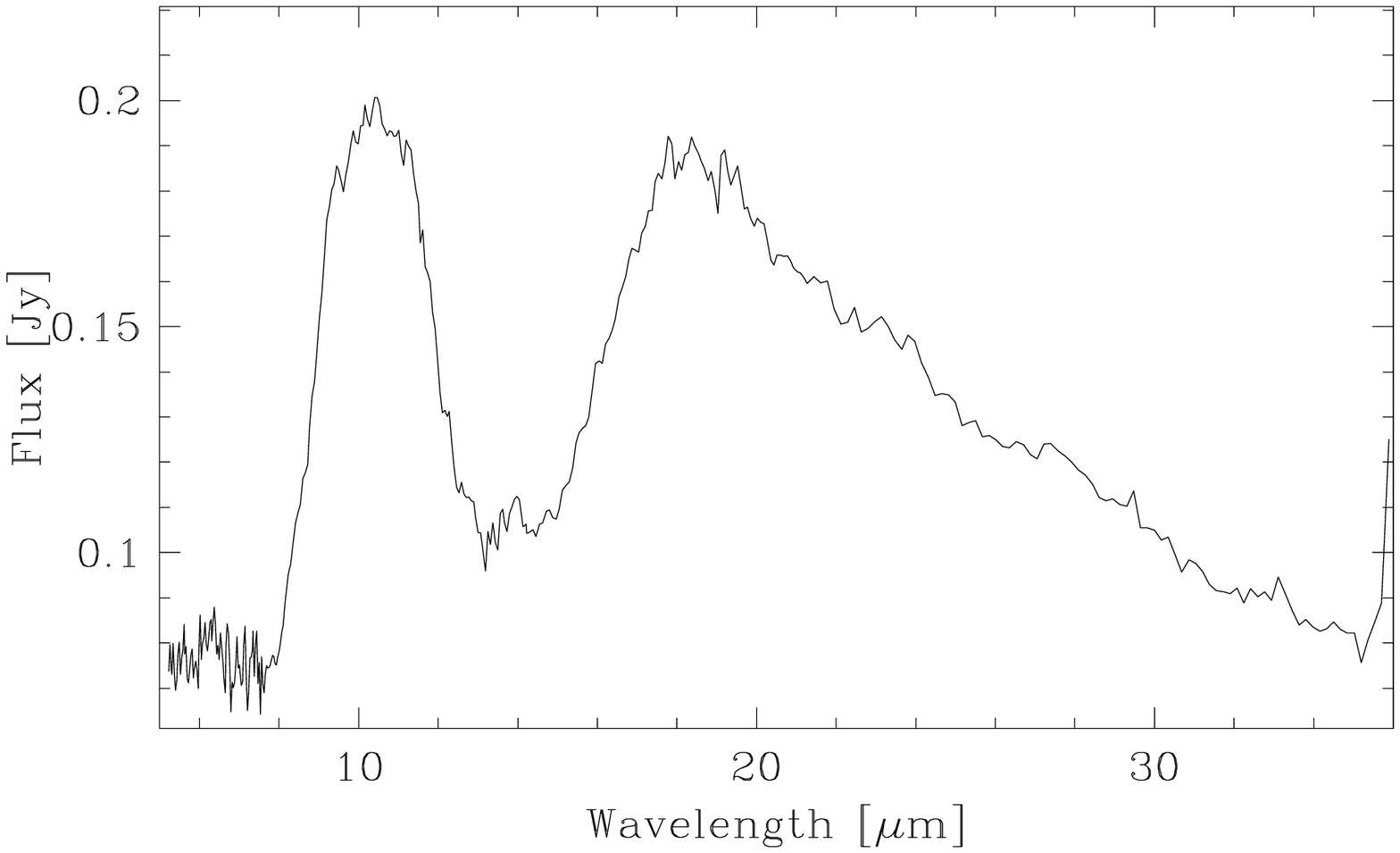}
\caption{\label{spectra4} Spectrum of the oxygen-rich AGB star 
IRAS 05003$-$6712 }
\end{figure}

\subsection{Bolometric magnitudes}
\label{mbol}

For a number of our targets, bolometric magnitudes were derived by
\citet{Wood1998} and \citet{Whitelock2003}. Table \ref{targets.dat} lists the
bolometric magnitudes derived from the {\it Spitzer} spectra and JHKs 2MASS
photometry. Where multi-epoch photometry was available
(Table~\ref{periods.dat}), those were used instead of the single-epoch 2MASS
data. The spectra were smoothed by a factor of 20. The J-band upper limit was
used as a real magnitude where necessary: this has no measurable effect on the
result (changing it by 1 mag affected the bolometric flux by less than 1 per
cent).  Zero flux was assumed at frequencies of 0\,Hz and $3\,10^{16}\,$Hz.  A
fourth-order polynomial was fitted (using IDL) to the spectra and data points
and the total flux was obtained by integration under the fit.  For the blue
carbon stars, an uncertainty is introduced by the lack of data points between
2 and 5$\mu$m.  Variability is the main limitation to the accuracy: the K-band
magnitude can vary by up to 1--2 mag in extreme cases \citep{LeBertre1992,
  Wood1998, Whitelock2003}. The result was converted to absolute magnitudes
assuming a distance modulus of 18.5 \citep[e.g.,][]{Alves2002}.  The distance
modulus varies by approximately 0.1 magnitude over the face of the bar
\citep{Lah2005} but we did not correct for such geometric effects.

\begin{figure}
\includegraphics[width=8cm]{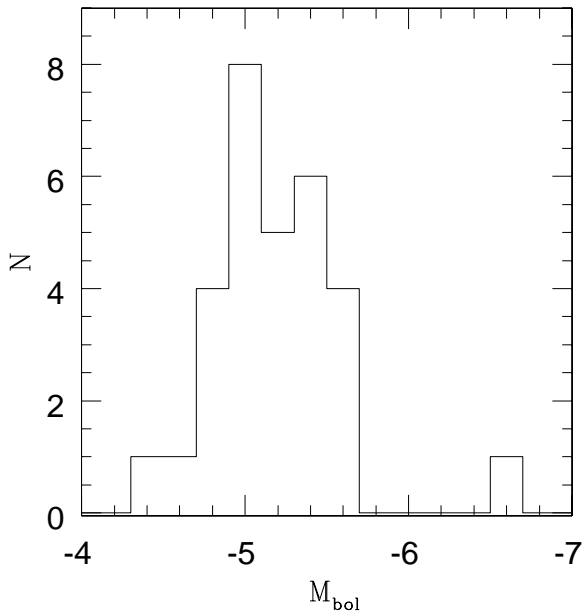}
\caption{\label{hist} Distribution of bolometric magnitudes }
\end{figure}

The histogram of the bolometric magnitudes is shown in Fig. \ref{hist}.

\section{Emission and absorption bands}

\subsection{Dust bands}
\label{dust}

Carbon stars show two major dust bands in this spectral region, at 11.3 and
30\,$\mu$m. Both features are seen in the majority of the current sample,
although not in all. Two other spectral features have been suggested as being
due to dust, but below we argue against this possibility.

\subsubsection{SiC dust emission}  The well-known 11.3-$\mu$m band is due to 
solid $\alpha$-SiC \citep[e.g.][]{Papoular1998, Clement2003}. The central
wavelength shows some variability in Galactic carbon stars, ranging between
11.15 and 11.7\,$\mu$m.  Longer wavelengths are correlated with lower dust
(continuum) temperatures \citep{Baron1987}. \citet{Clement2003} find that the
wavelength shift to the red can be reproduced in SiC particle conglomerates,
and they suggest that conglomerates form at higher density in the wind. In
this case the correlation with the dust temperature is an indirect one, caused
by a larger optical depth in the dust shell. SiC absorption components are
discussed by \citet{Speck2005}.

\subsubsection{MgS dust emission}
The 30-$\mu$m resonance is commonly seen in carbon-rich environments, in AGB
stars, post-AGB stars and planetary nebulae \citep{Forrest1981, Hony2002a}. It
consists of two subpeaks, at 26 and 33\,$\mu$m \citep{Volk2002}. The two peaks
always appear together, arguing for an origin in the same mineral.  The
resonance is attributed to MgS \citep{Goebel1985} which forms via
grain-surface reaction between Mg and H$_2$S \citep{Nuth1985}; the two peaks
have been attributed to different grain shapes \citep{Hony2002b}. A
contribution from CaS to the features cannot be fully excluded, but the
absence of any feature in oxygen-rich environments argues against this
\citep{Goebel1985}. The peak wavelength of the MgS feature shifts with
temperature, which makes it an important temperature diagnostic
\citep{Hony2004}. \citet{Hony2002b} argue that solid MgS may survive into the
ISM.

\subsubsection{Doubtful features} Two objects show indications of a 
broad, weak hump between 16 and 23\,$\mu$m, with unclear identification.  It
is most noticeable in the symbiotic C star LMC MSX 663, and may be seen
weakly in MSX LMC 754, a carbon star with hot dust. Neither object shows
evidence for MgS. The feature is likely an artifact caused by the broad
molecular C$_2$H$_2$ absorption band to the blue.

A possible emission feature at 8.6\,$\mu$m has been described in IRAS LRS (Low
Resolution Spectrometer) spectra of Galactic carbon stars \citep{Willems1988,
  Sloan1998}.  The feature correlates in strength with the SiC band and the
3\,$\mu$m band.  However, its reality is disputed: strong absorption at
7.5\,$\mu$m \citep{Aoki1999} together with 10\,$\mu$m absorption can mimic an
emission band. The spectra in our LMC sample show strong absorption at
7.5\,$\mu$m and absorption at 10\,$\mu$m, making it more likely that the
8.6\,$\mu$m peak is simply the continuum between two molecular bands, which we
will assume below.

\subsubsection{10 micron absorption} This absorption feature, which has 
been attributed to a dust component, is discussed in Section \ref{cthree}.

\subsection{Molecular bands}

The molecular bands are all at the short wavelength range of the IRS spectra.
\citet{Gautschy2004} discuss the contribution of different molecules to the
5--20-$\mu$m wavelength range. Example models for individual bands were
calculated by \citet{Jorgensen2000}: their models are shown at the of
Figs. \ref{spectra1} and \ref{spectra2}

\subsubsection{13.7 micron}
The narrow 13.7-$\mu$m band is prominent in most of the spectra.  This band is
due to the Q-branch C$_2$H$_2$ $\nu_5$ transition \citep{Tsuji1984, Aoki1999,
  Cernicharo1999}.  The narrow band is sitting in the much broader P,R
branches, giving rise to an absorption feature extending from 12 to 16-$\mu$m,
which is best seen in the spectrum of IRAS 05190$-$6748. The 11.3-$\mu$m SiC
emission feature makes it difficult to judge how strong this broad absorption
band is.  Models of Galactic carbon stars \citep{Gautschy2004} predict a much
stronger broad band than is observed. This can be explained if the mass-losing
envelope adds a molecular emission feature. The sharp 13.7-$\mu$m band is very
strong in our spectra, while in Galactic stars it has been dubbed the 'weak'
feature \citep{Gautschy2004}. The sharp feature originates from colder gas,
while hot (up to photospheric temperatures) gas mainly shows the broad band
\citep{Jorgensen2000}.

\subsubsection{14.3 micron}
The 14.3-$\mu$m HCN band seen in Galactic carbon stars \citep{Aoki1999,
  Cernicharo1999} is weak in our sample.  It is probably present in the
spectrum of TRM 72. Otherwise the feature is not clear, and is certainly much
weaker than the C$_2$H$_2$ band.  This is similar to the situation in Galactic
stars, where its weakness is attributed to in-fill by an emission component
\citep{Aoki1999}.

\subsubsection{7.5 micron}
A noticeable absorption band is present at 7.5\,$\mu$m. Galactic carbon stars
have contributions in this wavelength region from C$_2$H$_2$, HCN and CS
\citep{Goebel1981, Aoki1998}. The sample of \citet{Aoki1998} consists of blue
carbon stars, which differ in effective temperature from our sample.

The shape of the two shallow features found in our spectra resembles
C$_2$H$_2$. The diatomic molecule CS shows a band head which we do not see,
and the HCN band is located at shorter wavelength, typically centered around
7.0\,$\mu$m (although this central wavelength depends on the excitation
temperature).  Comparison of our LMC spectra and Galactic spectra indicates a
difference in the shape of the 7.5-$\mu$m band, supporting different origins
for Galactic and LMC carbon stars. \citet{Matsuura2002a, Matsuura2005a} show
that at low metallicity, C$_2$H$_2$ becomes more abundant compared to HCN.
Following this, we identify the observed band for LMC stars with C$_2$H$_2$.


\subsubsection{5 micron}
At the blue edge of the spectra the spectral energy distributions drop
sharply, due to a strong absorption band which extends beyond the blue limit
of the wavelength range. There are two molecules which contribute at this
wavelength in carbon-rich environments: CO and C$_3$ \citep{Jorgensen2000}.
CO is of photospheric origin and is present even in K stars. Its presence in
e.g. the unobscured source GRRV\,38 does not require a circumstellar shell.
C$_3$ also is photospheric in carbon stars.  The two bands differ at the red
end: C$_3$ ends at 6\,$\mu$m, while CO tapers off slowly up to 8\,$\mu$m.

SiS has three bands at 6.6--6.7\,$\mu$m but the IRS resolution is not high
enough to detect these; $R=2000$ would be required.

\subsection{The 10-micron band}
\label{cthree}

An absorption feature at 10\,$\mu$m may be present in most stars. It is seen
only occasionally in Galactic carbon stars \citep{Volk2000, Sloan1998}.
\citet{Jorgensen2000} find it in R Scl but not in three other carbon stars
they study.  The origin is disputed.  \citet{Clements2005} fit this band using
solid silicon-nitrite particles, Si$_3$N$_4$. \citet{Speck1997} and
\citet{Volk2000} suggest interstellar silicate absorption; to fit the shape,
additional SiC absorption is required.  \citet{Jorgensen2000} identify the
band instead with C$_3$ (see their Fig. 11).

The presence of the band in our LMC stars argues against the interstellar
origin proposed by \citet{Speck1997}, because the line-of-sight extinction to
the LMC is too small to give significant silicate absorption. A contribution
from SiC absorption, as required by the silicate hypothesis, is also ruled
out: this would only be in absorption in dense, cool shells, while we see the
10-$\mu$m absorption even in blue, unobscured stars.

This last point, the presence of the feature in all stars, argues against a
dust source and bring into question the suggested identification
\citep{Clements2005} with silicon-nitrite.  A molecular origin is favoured, in
particular a molecule with a photospheric origin.

We finally note the prevalence of the feature in our sample, as compared to
Galactic stars. This points towards a molecule which is more abundant in the
LMC (at low metallicity) than in the Galaxy. \citet{Matsuura2002a} have shown
that this is the case for carbon-bearing molecules, due to a higher C/O ratio.
The combined effect, at lower metallicity, of increased efficiency of third
dredge-up of carbon \citep{Wood1981} and lower oxygen abundance leads to
higher C/O ratios at lower metallicity.

The identification suggested by \citet{Jorgensen2000} fits our available
constraints. We therefore favour C$_3$ as the probable carrier. However,
confirmation will require a laboratory measurement of its 10-$\mu$m spectrum.

\subsection{Band strengths}
\label{band_strengths}

\begin{table}
\caption[]{\label{cont.def} Wavelengths used to estimate the continua for the
  SiC and C$_2$H$_2$ spectral features.}
\begin{flushleft}
\begin{tabular}{lllllll}
\hline
Features & $\lambda$ [$\mu$m] & Blue continuum [$\mu$m]& Red continuum [$\mu$m]\\
\hline
C$_2$H$_2$ &   7.5  & 6.08-6.77    & 8.22-8.55 \\
SiC        &   11.3 & 9.50-10.10   & 12.80-13.40 \\
C$_2$H$_2$ &   13.7 & 12.80-13.40  & 14.10-14.70 \\

\hline \\
\end{tabular}
\end{flushleft}
\end{table}

\begin{figure}
\includegraphics[width=7cm]{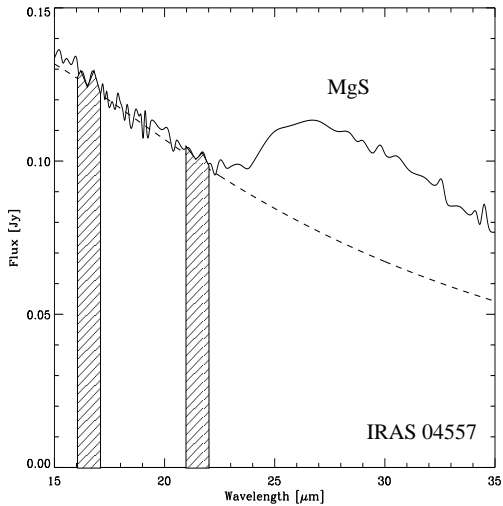}
\caption{\label{mgs_iras04557.eps} The continuum underlying the MgS feature is
  subtracted using a black body, whose temperature is defined using the
  [16.5]$-$[21.5] color. The hatched bars display the wavelengths used to
  determine the [16.5] and [21.5] colors. The dashed line represents the
  black body fit to the spectrum of IRAS 04557$-$6753 taken as an example.}
\end{figure}

We define the strength of the molecular absorption bands through their
equivalent widths, and quantify the dust emission bands by their integrated
line-to-continuum ratio.  We use the adjacent continuum to define a linear
(sloping) continuum distribution across each band and define the equivalent
width in the usual way. The wavelenghts used to estimate the continuum for the
C$_2$H$_2$ (7.5\,$\mu$m), SiC (11.3\,$\mu$m) and C$_2$H$_2$ (13.7\,$\mu$m) are
listed in Table~\ref{cont.def}. For the 13.7-$\mu$m band, the equivalent width
mainly refers to the narrow component: the broad component extends beyond the
chosen continuum wavelengths.

The continuum underneath the MgS band cannot be measured in this way, because
the band is too broad and extends beyond the red edge of the spectral coverage
of the IRS. We therefore instead use a black body to approximate the
continuum, and give the strength as a line-to-continuum ratio integrated over
the feature up to 38\,$\mu$m. The black body is defined using the
[16.5]$-$[21.5] colour discussed below (see Fig.~\ref{mgs_iras04557.eps}).
Table \ref{ew.dat} lists the resulting values.



\begin{table*}
\caption[]{\label{ew.dat} Strength of the molecular and dust features, in
terms of either the equivalent width (EW) for molecular absorption bands, and
the line-to-continuum ratio (L/C) for the dust emission bands. The central
wavelength of the SiC band is also listed. The final column gives the
continuum (black-body) temperature, derived from the [16.5]$-$[21.5] colour
listed in Table \ref{colors.dat} }
\begin{flushleft}
\begin{tabular}{llllllllllllllll}
\hline
target         &  EW (7.5\,$\mu$m)     & EW (13.7\,$\mu$m)    & L/C (SiC)       & $\lambda_c$  & L/C (MgS)  & T(K)  \\
\hline
TRM 72          &  0.052$\pm$0.005 & 0.068$\pm$0.006 & 0.073$\pm$0.002 & 11.37 & 0        & 1018 \\
IRAS 04557      &  0.156$\pm$0.004 & 0.066$\pm$0.003 & 0.123$\pm$0.003 & 11.30 & 0.412     & 426  \\
IRAS 05009      &  0.069$\pm$0.005 & 0.062$\pm$0.004 & 0.159$\pm$0.003 & 11.31 & 0.249     & 619  \\
IRAS 05112      &  0.121$\pm$0.005 & 0.066$\pm$0.003 & 0.099$\pm$0.002 & 11.34 & 0.271     & 396  \\
IRAS 05113      &  0.133$\pm$0.008 & 0.049$\pm$0.004 & 0.094$\pm$0.004 & 11.34 & 0.215     & 540  \\
IRAS 05132      &  0.092$\pm$0.006 & 0.071$\pm$0.005 & 0.117$\pm$0.004 & 11.27 & 0.198     & 557  \\
IRAS 05190      &  0.135$\pm$0.004 & 0.074$\pm$0.004 & 0.104$\pm$0.003 & 11.37 & 0.374     & 398  \\
IRAS 05278      &  0.082$\pm$0.004 & 0.075$\pm$0.003 & 0.111$\pm$0.003 & 11.32 & 0.407     & 421  \\
GRRV 38         &  0.218$\pm$0.023 & 0.118$\pm$0.025 & 0.143$\pm$0.019 & 11.14 & 0        & 858  \\
IRAS 05295      &  0.092$\pm$0.007 & 0.057$\pm$0.004 & 0.086$\pm$0.002 & 11.34 & 0.162     & 462  \\
IRAS 05360      &  0.092$\pm$0.004 & 0.051$\pm$0.006 & 0.180$\pm$0.006 & 11.19 & 0.624     & 395  \\
MSX LMC 219      &  0.111$\pm$0.005 & 0.061$\pm$0.004 & 0.108$\pm$0.003 & 11.29 & 0.336     & 393  \\
MSX LMC 341      &  0.083$\pm$0.007 & 0.063$\pm$0.005 & 0.153$\pm$0.004 & 11.16 & 0.637     & 482  \\
MSX LMC 349      &  0.079$\pm$0.007 & 0.049$\pm$0.005 & 0.173$\pm$0.004 & 11.18 & 0.713     & 470  \\
MSX LMC 441      &  0.129$\pm$0.007 & 0.056$\pm$0.004 & 0.125$\pm$0.003 & 11.26 & 0.446     & 427  \\
MSX LMC 443      &  0.135$\pm$0.008 & 0.054$\pm$0.007 & 0.184$\pm$0.006 & 11.18 & 0.459     & 467  \\
MSX LMC 494      &  0.109$\pm$0.007 & 0.039$\pm$0.003 & 0.080$\pm$0.003 & 11.31 & 0        & 780  \\
MSX LMC 601      &  0.016$\pm$0.005 & 0.060$\pm$0.005 & 0.056$\pm$0.003 & 11.30 & 0        & 703  \\
MSX LMC 679      &  0.122$\pm$0.006 & 0.044$\pm$0.004 & 0.151$\pm$0.004 & 11.22 & 0.477     & 415  \\
MSX LMC 743      &  0.158$\pm$0.005 & 0.042$\pm$0.004 & 0.156$\pm$0.004 & 11.24 & 0.383     & 455  \\
MSX LMC 749      &  0.120$\pm$0.003 & 0.076$\pm$0.003 & 0.058$\pm$0.004 & 11.25 & 0        & 865  \\
MSX LMC 754      &  0.121$\pm$0.006 & 0.053$\pm$0.004 & 0.146$\pm$0.006 & 11.32 & 0        & 726  \\
MSX LMC 967      &  0.085$\pm$0.003 & 0.058$\pm$0.003 & 0.086$\pm$0.003 & 11.30 & 0        & 742  \\
NGC 1978 MIR1    &  0.157$\pm$0.004 & 0.064$\pm$0.006 & 0.162$\pm$0.005 & 11.30 & 0.528     & 446  \\
NGC 1978 IR1     &  0.123$\pm$0.003 & 0.070$\pm$0.004 & 0.094$\pm$0.004 & 11.23 & 0        & 1179 \\
NGC 1978 IR4     &  0.172$\pm$0.010 & 0.082$\pm$0.009 & 0.107$\pm$0.006 & 11.50 & 0        & \llap{$>$}$10^3$ \\
TRM 88          &  0.051$\pm$0.006 & 0.064$\pm$0.005 & 0.065$\pm$0.004 & 11.26 & 0        & 684  \\
\hline \\
\end{tabular}
\end{flushleft}
\end{table*}

\section{Continuum definition: The Manchester system}

\begin{table}
\caption[]{\label{colours.def} Mid-infrared 
continuum bands for carbon stars for the so-called 'Manchester system'. 
The last column
gives the adopted flux density corresponding to zero magnitude.
  }
\begin{flushleft}
\begin{tabular}{llllll}
\hline
central $\lambda$ [$\mu$m] & $\lambda$-range [$\mu$m]  & $F_0$ [Jy] \\
\hline
   6.4 & 6.25--6.55 & 96.5 \\
   9.3 & 9.1--9.5   & 45.7 \\
   16.5 & 16--17    & 15.4 \\
   21.5 & 21--22   & 9.1 \\
\hline \\
\end{tabular}
\end{flushleft}
\end{table}

\begin{figure}
\includegraphics[width=7cm]{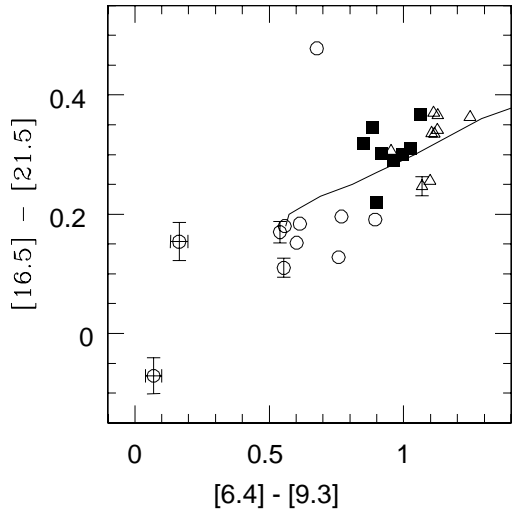}
\caption{\label{spitcol.ps} The [6.4]$-$[9.3] versus [16.5]$-$[21.5]
  colour--colour diagram. 
  Symbols are defined in Fig. \ref{k_jk.ps}. The drawn line represents
 a series of Dusty models, shifted down by 0.15\,mag in [16.5]$-$[21.5]. }
\end{figure}

Much of the observed spectral range is covered by the molecular and dust bands
discussed above. This leaves the continuum poorly defined: one needs to select
specific wavelength regions which avoid these bands. The broadband IRAS
filters are clearly not a good approximation. The blue continuum is most
affected by molecular bands; the red continuum is in many cases obliterated
beyond 15\,$\mu$m by the 30-$\mu$m MgS band.

We have selected four narrow regions which can be used for dusty carbon stars
(not for oxygen-rich stars!). The regions avoid most, but not all, of the
obvious absorption and emission bands.  The continuum bands are defined in
Table~\ref{colours.def}.  The bands will be called the Manchester system.
Table~\ref{colours.def} also lists the adopted zeropoint for the bands. These
are extrapolated from the standard bands, and aim to return a zero colour for
a Rayleigh-Jeans tail.

The 6.4-$\mu$m band is situated between the SiS bands at 6.6\,$\mu$m and CS at
7\,$\mu$m on the red side, and the CO band at 5\,$\mu$m on the blue
\citep{Aoki1998}. The wing of the CO feature may still affect the chosen band.
There is also overlap with the PAH 6.2-$\mu$m band but this band requires UV
excitation and is seen in planetary nebulae and post-AGB stars but not in AGB
star spectra.

The 9.3-$\mu$m band is blueward of the SiC dust band, and red of the CS bands
and C$_2$H$_2$ bands at 7--8\,$\mu$m \citep{Aoki1999}. There are also bands
from molecules such as NH$_3$ and C$_2$H$_4$ which may depress the continuum
by a few per cent \citep{Cernicharo2001a}. However, it is situated in the
C$_3$ band discussed above, which is the main uncertainty in this spectral
region and is significant for spectra dominated by photospheric emission. For
these stars, the 9.3-$\mu$m band will underestimate the true continuum.

The two bands at 16.5 and 21.5\,$\mu$m do not overlap with any strong
molecular or dust features in carbon-rich material: it is situated just
redward of the broad P-branch 14-micron C$_2$H$_2$ band, which extends to
16\,$\mu$m. There is a weak hydrocarbon feature at the blue edge of the chosen
band \citep{Cernicharo2001b}. The strong 21-$\mu$m dust feature which is still
unidentified is seen in post-AGB stars only. The Manchester system should
therefore be used for carbon-rich AGB stars only.  The regions also avoid
potential instrumental problems in the IRS, such as the LL1/LL2 interface.

The continuum fluxes of all observed targets in the Manchester system are
listed in Table \ref{colors.dat}. These are derived by integrating the IRS
spectra over the bands.  We give the colours using the zeropoints of Table
\ref{colours.def}.  Similar colours for SMC stars are given in
\citet{Sloan2006}.  Table \ref{colors.dat} lists a black-body colour
temperature derived from the [16.5]$-$[21.5] colour.  For the bluest spectra,
the values become poorly defined as the red colour approaches the
Rayleigh-Jeans limit, but for the majority of objects a reasonable value is
derived. Apart from NGC\,1978\,IR1  with photospheric colours, the colour
temperatures range from $\sim$ 1200\,K for NGC1978 IR1 to around 400\,K for
the coolest objects.

\begin{table*}
\caption[]{\label{colors.dat} Photometry: fluxes and colours using
the Manchester  narrow  continuum bands for carbon stars. For the
colours in the last two columns, we adopt the zeropoints
of Table \ref{colours.def} 
  }
\begin{flushleft}
\begin{tabular}{llllllllllllllll}
\hline
target         &  $F_{6.4}$  & $F_{9.3}$  & $F_{16.5}$  &
$F_{21.5}$  & [6.4]$-$[9.5] & [16.5]$-$[21.5] \\
               & [Jy]             & [Jy]            & [Jy]            & [Jy]
& [mag] & [mag] \\
\hline
TRM 72           &  0.211$\pm$0.001 & 0.201$\pm$0.000 & 0.119$\pm$0.001 & 0.079$\pm$0.001 &  0.758$\pm$0.005 &  0.128$\pm$0.015 \\
IRAS 04557      &  0.133$\pm$0.001 & 0.174$\pm$0.001 & 0.128$\pm$0.001 & 0.102$\pm$0.001 &  1.104$\pm$0.010 &  0.336$\pm$0.009 \\
IRAS 05009      &  0.228$\pm$0.001 & 0.247$\pm$0.001 & 0.156$\pm$0.001 & 0.112$\pm$0.001 &  0.898$\pm$0.005 &  0.220$\pm$0.013 \\
IRAS 05112      &  0.231$\pm$0.002 & 0.309$\pm$0.001 & 0.233$\pm$0.001 & 0.192$\pm$0.001 &  1.127$\pm$0.008 &  0.336$\pm$0.007 \\
IRAS 05113      &  0.160$\pm$0.001 & 0.209$\pm$0.001 & 0.154$\pm$0.001 & 0.115$\pm$0.001 &  1.098$\pm$0.008 &  0.256$\pm$0.011 \\
IRAS 05132      &  0.143$\pm$0.001 & 0.181$\pm$0.001 & 0.132$\pm$0.001 & 0.098$\pm$0.001 &  1.068$\pm$0.008 &  0.247$\pm$0.016 \\
IRAS 05190      &  0.225$\pm$0.002 & 0.336$\pm$0.000 & 0.295$\pm$0.001 & 0.242$\pm$0.001 &  1.247$\pm$0.010 &  0.363$\pm$0.007 \\
IRAS 05278      &  0.604$\pm$0.003 & 0.807$\pm$0.002 & 0.563$\pm$0.004 & 0.454$\pm$0.002 &  1.126$\pm$0.006 &  0.341$\pm$0.008 \\
GRRV 38         &  0.067$\pm$0.001 & 0.037$\pm$0.001 & 0.021$\pm$0.001 & 0.014$\pm$0.000 &  0.165$\pm$0.033 &  0.154$\pm$0.032 \\
IRAS 05295      &  0.150$\pm$0.001 & 0.171$\pm$0.000 & 0.115$\pm$0.001 & 0.090$\pm$0.001 &  0.953$\pm$0.006 &  0.306$\pm$0.009 \\
IRAS 05360      &  0.107$\pm$0.001 & 0.135$\pm$0.001 & 0.089$\pm$0.001 & 0.074$\pm$0.001 &  1.062$\pm$0.006 &  0.367$\pm$0.014 \\
MSX LMC 219      &  0.112$\pm$0.001 & 0.148$\pm$0.001 & 0.097$\pm$0.000 & 0.080$\pm$0.001 &  1.112$\pm$0.009 &  0.370$\pm$0.009 \\
MSX LMC 341      &  0.109$\pm$0.000 & 0.126$\pm$0.001 & 0.078$\pm$0.001 & 0.060$\pm$0.000 &  0.963$\pm$0.006 &  0.291$\pm$0.012 \\
MSX LMC 349      &  0.141$\pm$0.001 & 0.166$\pm$0.000 & 0.103$\pm$0.001 & 0.080$\pm$0.001 &  0.994$\pm$0.007 &  0.300$\pm$0.009 \\
MSX LMC 441      &  0.137$\pm$0.001 & 0.181$\pm$0.001 & 0.119$\pm$0.001 & 0.096$\pm$0.000 &  1.115$\pm$0.013 &  0.335$\pm$0.008 \\
MSX LMC 443      &  0.083$\pm$0.001 & 0.092$\pm$0.000 & 0.055$\pm$0.000 & 0.042$\pm$0.000 &  0.918$\pm$0.008 &  0.302$\pm$0.014 \\
MSX LMC 494      &  0.094$\pm$0.000 & 0.073$\pm$0.000 & 0.035$\pm$0.000 & 0.024$\pm$0.000 &  0.540$\pm$0.006 &  0.170$\pm$0.018 \\
MSX LMC 601      &  0.115$\pm$0.001 & 0.124$\pm$0.000 & 0.074$\pm$0.000 & 0.052$\pm$0.001 &  0.894$\pm$0.006 &  0.191$\pm$0.012 \\
MSX LMC 663      &  0.116$\pm$0.001 & 0.102$\pm$0.000 & 0.078$\pm$0.000 & 0.072$\pm$0.000 &  0.677$\pm$0.006 &  0.478$\pm$0.007 \\
MSX LMC 679      &  0.135$\pm$0.001 & 0.144$\pm$0.000 & 0.085$\pm$0.001 & 0.069$\pm$0.000 &  0.886$\pm$0.006 &  0.346$\pm$0.010 \\
MSX LMC 743      &  0.093$\pm$0.000 & 0.113$\pm$0.000 & 0.067$\pm$0.000 & 0.053$\pm$0.000 &  1.026$\pm$0.006 &  0.311$\pm$0.009 \\
MSX LMC 749      &  0.238$\pm$0.001 & 0.196$\pm$0.000 & 0.101$\pm$0.000 & 0.068$\pm$0.001 &  0.602$\pm$0.003 &  0.152$\pm$0.011 \\
MSX LMC 754      &  0.086$\pm$0.000 & 0.072$\pm$0.000 & 0.033$\pm$0.000 & 0.023$\pm$0.000 &  0.614$\pm$0.009 &  0.184$\pm$0.014 \\
MSX LMC 967      &  0.182$\pm$0.000 & 0.144$\pm$0.001 & 0.074$\pm$0.001 & 0.051$\pm$0.000 &  0.558$\pm$0.005 &  0.180$\pm$0.011 \\
NGC 1978 MIR1    &  0.112$\pm$0.001 & 0.116$\pm$0.000 & 0.072$\pm$0.001 & 0.057$\pm$0.000 &  0.852$\pm$0.006 &  0.318$\pm$0.009 \\
NGC 1978 IR4  &  0.042$\pm$0.000 & 0.021$\pm$0.000 & 0.012$\pm$0.000 & 0.007$\pm$0.000 &  0.079$\pm$0.024 &  0.008$\pm$0.031 \\
NGC 1978 IR2     &  0.045$\pm$0.001 & 0.022$\pm$0.000 & 0.012$\pm$0.000 & 0.006$\pm$0.000 &  0.061$\pm$0.025 &\llap{$-$}0.134$\pm$0.026 \\
NGC 1978 IR1     &  0.106$\pm$0.000 & 0.084$\pm$0.001 & 0.039$\pm$0.000 & 0.025$\pm$0.000 &  0.554$\pm$0.007 &  0.110$\pm$0.016 \\
TRM 88          &  0.098$\pm$0.001 & 0.094$\pm$0.001 & 0.052$\pm$0.001 & 0.037$\pm$0.000 &  0.769$\pm$0.009 &  0.196$\pm$0.013 \\
\hline \\
\end{tabular}
\end{flushleft}
\end{table*}

\section{Continuum colours and optical depth}

\subsection{Manchester colours}
The IRS colour--colour diagram based on the Manchester bands for our targets
is shown in Fig. \ref{spitcol.ps}. Error bars are indicated where the size of
the bars exceeds the size of the symbol. We note that the errors are
calculated from the tabulated noise on the spectrum, and do not include
systematic effects such as calibration uncertainties or background
subtraction.  The majority of the stars fall on a well determined sequence.
There three exceptions: two stars are much bluer in [6.4]$-$[9.3], and one
star is too red in [16.5]$-$[21.5] by approximately 0.25\,mag.

The first two stars are GRRV\,38 and NGC\,1978\,IR4.  Both of these have
spectral energy distributions largely consistent with naked carbon stars. They
are also faint in the IRS wavelength range and the colours have larger
uncertainties than for the other stars, but the errorbars shown on the figure
indicate that the flux uncertainties do not affect their positions in this
diagram. The last star is MSX\,LMC\,663, in which the [21.5] band includes an
additional emission component.

All other stars show continuum emission dominated by circumstellar dust over
all wavelenghts considered here. The good correlation between the two colours
confirms the choice of the continuum regions.

\subsection{Comparison models}
\label{cont_models}
We calculated a series of benchmark dust models, to compare the colours. The
models used the DUSTY code, with a density distribution of a radiation-driven
wind \citep{Elitzur2001}, $\rho \sim \rho^{-1.8}$. Models were run with an
optical depth $\tau$ at 1\,$\mu$m ranging from 0.01 to 100. For the central
star we assumed a black body of $T = 2800\,$K.  The inner radius of the shell
was set at a dust temperature of 1000\,K. A dust mixture of 95 per cent
amorphous carbon and 5 per cent SiC was used. The resultant spectra were used
to measure the flux through the bandpasses, converted to colours using the
zeropoints of Table \ref{colours.def}. Such models lack the important
molecular bands and are not meant to fit the observed spectra: this will be
the subject of a forthcoming paper.

The models are systematically too red in [16.5]$-$[21.5] by about 0.15\,mag.
The drawn line in Fig. \ref{spitcol.ps} shows the models, where we shifted the
model tracks down by 0.15\,mag to compensate for the offset. The cause of the
offset is not known.

The models for $\tau=0.01$ yield a colour of [6.4]$-$[9.3]$= 0.56$. At these
wavelengths, a 2800\,K black body is considerably redder than a Rayleigh-Jeans
curve.  The observed cutoff in this colour (Fig. \ref{spitcol.ps}) is
consistent with the models, but the two stars with photospheric spectra are
considerably bluer. This blueshift reflects the effect of photospheric
molecular bands, such as suppression of the [9.3] flux by the proposed C$_3$
band.  In support of this supression by C$_3$, the [6.4]$-$[9.3] colour of the
symbiotic C star MSX LMC 663, which also has little dust but different 
photospheric molecular bands, is in good agreement with the models.

We conclude that [6.4]$-$[9.3]$ <0.5$ is indicative of a naked carbon star.
The implication is that the Manchester [9.3] band may not define a stellar
continuum for a naked carbon star, but it can be used instead to judge the
strength of the absorption band.

\subsection{Optical depth}

\begin{figure}
\includegraphics[width=8.5cm]{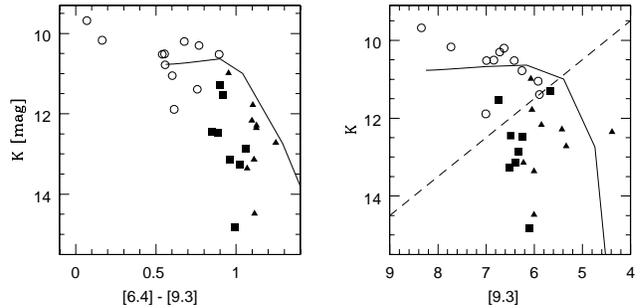}
\caption{\label{Kext.ps} The K magnitude versus the [6.4]$-$[9.3] colour
and versus the [9.3] magnitude.
  Symbols are as defined in Fig. \ref{k_jk.ps}: squares and triangles
indicate stars with both SiC and MgS. The drawn lines refer to the 
benchmark models of Section \ref{cont_models}.}
\end{figure}

High optical depth in a circumstellar shell affects the spectral energy
distribution in two ways.  Extinction within the wavelength of interest
suppresses the bluest bands, and absorption of optical and near-infrared
stellar emission lowers the dust temperature further out.\footnote{In
  individual cases, density variations which differ from an $r^{-2}$
  distribution can lead to dust temperatures which mimic optical-depth
  effects.}  This is illustrated for our sample in Fig.  \ref{Kext.ps}.

The left panel shows the K-magnitude versus the [6.4]$-$[9.3] colour.  The
relation between redder colour and fainter K-band magnitude in Fig.
\ref{Kext.ps} confirms the effects of optical depth. The right panel shows the
K-band versus the [9.3] band magnitude: the latter is not affected by
extinction within the range of optical depths probed by our sample.  It is
also close to the peak of the dust spectral energy distribution, and varies
little as function of temperature.  Indeed, the distribution of [9.3]
magnitudes remains fairly narrow while the K-band magnitude fades. (One
outlier in this plot is IRAS\,05278$-$6942, which has a higher bolometric
luminosity.)  At brighter K, the [9.3] magnitude becomes a little fainter,
tracing the effect of less dust.

The drawn curves show the same models as discussed in section
\ref{cont_models}. For a single luminosity, it shows the expected behaviour
with increasing optical depth, excluding the effects of molecular bands.  As
the depth of the shell increases, first the [9.3] magnitude becomes brighter,
before the K-band begins to fade sharply beyond a critical optical depth. The
curves follow the general behaviour of the observed spectra, but we again note
that no attempt has been made to fit the data.

The optical depth is a measure of the mass-loss rate.  The models imply that
the K\,$-$\,[9.3] colour can be used as an indicator for the optical depth.
The dashed line in the right panel of Fig.\ref{Kext.ps} shows a constant
K\,$-$\,[9.3] corresponding to the onset of K-band fading. The filled symbols
(squares and triangles) show sources with 30-$\mu$m MgS emission. The dashed
line shows that the presence of especially this band is related to the optical
depth in the shell.


\section{Discussion}

\subsection{Carbon star colour selection}
\label{carbonclass}

Almost all sources presented here are carbon stars. Only two oxygen-rich
objects were found in our entire sample of LMC stars, and only one is
classified as an AGB star.

On one hand, this result is entirely expected. The stars at the beginning of
the mass-loss sequence shown in Fig \ref{colours.eps}(a) are already optical
carbon stars and it is expected that more evolved stars on the sequence would
also be C stars. Such an expectation would only break down for massive and
relatively rare AGB stars that lie above the the sequence of Fig.
\ref{colours.eps}(a) since such stars do not ever become carbon stars.

\begin{figure}
\includegraphics[width=8.5cm,clip=true]{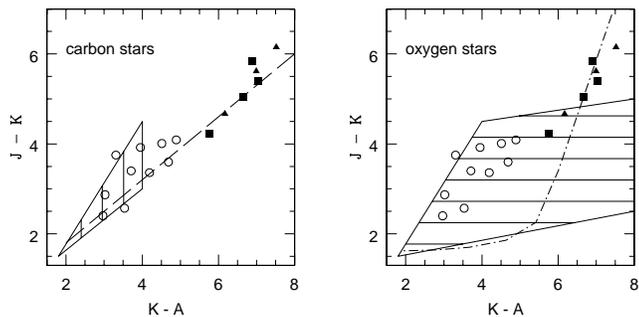}
\caption{\label{KmA.ps} Source classification diagram based on
  \citet{Egan2001}. The hashed area in the left panel indicates the colour
  range of model carbon stars, taken from the SKY model of
  \citet{Wainscoat1992}.  The hashed region in the right panel shows the same
  for oxygen-rich stars.  The symbols are our carbon-star sample. The dashed
  line in the left panel indicates the observed locus of obscured carbon
  stars. The dash-dotted line in the right panel shows the colour sequence for
  oxygen-rich stars, as derived from DUSTY models.}
\end{figure}

On the other hand, based on the colour classification of \citet{Egan2001}, a
higher fraction of oxygen-rich dust shells would be expected.  The colour
criterion to separate oxygen-rich and carbon-rich stars, suggested by
\citet{Egan2001}, is based on the J$-$K versus K$-$A colours. The diagram is
shown in Fig. \ref{KmA.ps}. The hashed regions indicate the colour range shown
by their model carbon stars (left panel) and oxygen stars (right panel), based
on the SKY model of \citet{Wainscoat1992}.  Overplotted are the location of
the stars in our sample. These mostly have colours redder than shown by any of
the model carbon stars included in the SKY model. The distribution of our
stars suggests that the SKY model can be extrapolated to redder stars.  For
obscured stars, there is considerable overlap in continuum colours between
oxygen-rich and carbon-rich stars.

There is however an effect from the silicate feature wich dominates the
10-$\mu$m spectrum of oxygen-rich stars.  The effect is shown for the IRAS
12-$\mu$m band by \citet{vanLoon1997}. At low mass-loss rates, the silicate is
strongly in emission and enhances the K$-$[12].  At higher mass-loss rates,
the silicate becomes a strong absorption feature which reduces the K$-$[12].
At a particular J$-$K, oxygen-rich stars with low mass-loss rates are redder
in K$-$[12] than carbon stars, whilst for high mass-loss rates the opposite is
true. Inevitably, there is a range where classification is not possible using
these colours.

The K$-$A colour shows a similar intrusion by silicate, but the MSX A-band
(6.8--10.8\,$\mu$m) contains more continuum and less silicate than does the
IRAS 12-$\mu$m band, and therefore the flux in the A-band does not decline as
much when silicate goes into absorption.  We calculated similar models as in
Section \ref{cont_models}, but for oxygen-rich dust. The stellar temperature
was taken as 3200\,K, and an \citet{Ossenkopf1992} warm 'circumstellar'
silicate model was assumed. (These silicate optical constants may not be
optimal for circumstellar dust.) The dashed line in the right panel of Fig.
\ref{KmA.ps} shows the resulting track with optical depth.  It does indeed
traverse the region used by \citet{Egan2001} but the K$-$A colour reaches an
effective limit whilst the J$-$K colour continues to redden.  The accuracy of
this track is limited.  The carbon-rich models of Section \ref{cont_models}
could not be used, because the molecular bands are rather important for the
stellar and circumstellar spectra at these wavelengths, and these bands are
missing from the DUSTY models.  Instead we draw a line through the observed
data points, which are known to be true locations of carbon stars.  At the
highest mass-loss rates, this track is likely to steepen considerably and to
run close to the model-track for oxygen-rich stars. At low mass-loss rate
C and O-rich stars are well-separated, but not at high $\dot M$.

\begin{figure*}
\includegraphics[width=17cm,clip=true]{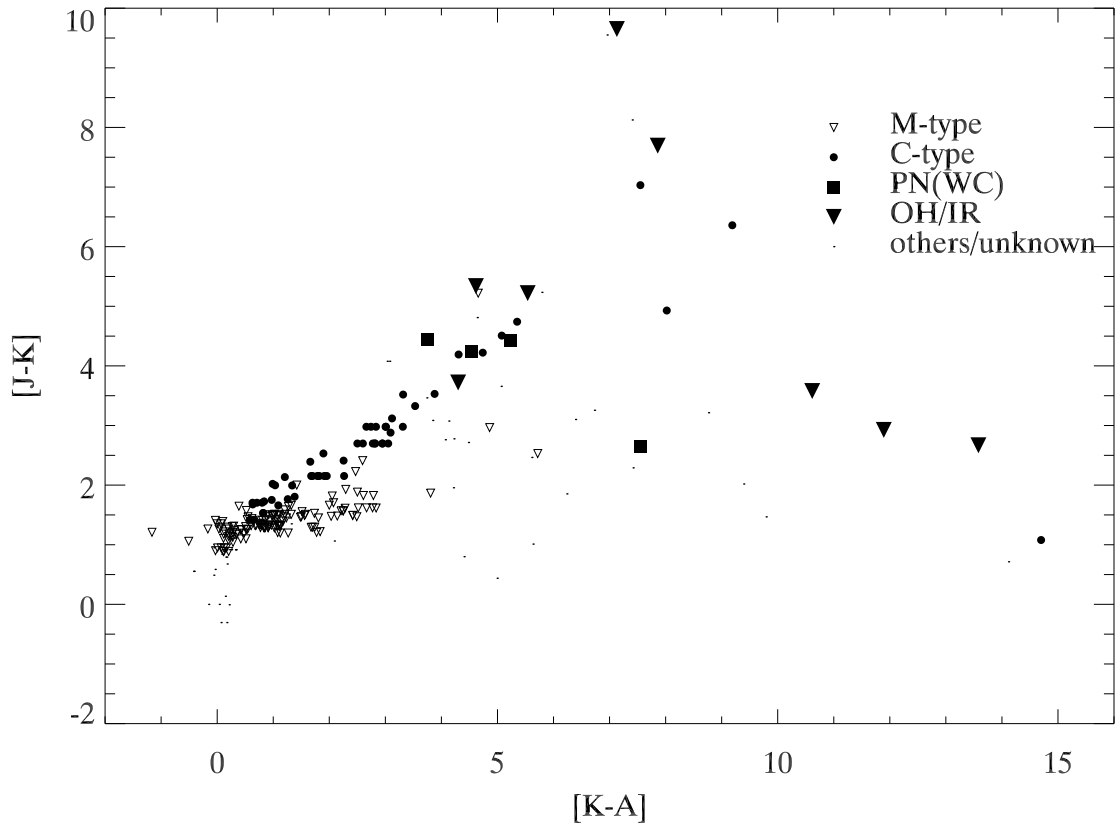}
\caption{\label{iso.ps} The J$-$K vs K$-$A diagram for evolved stars
observed by ISO}
\end{figure*}
 
We used ISO/SWS and 2MASS data of Galactic stars to further examine the
usefulness of nature of the J$-$K vs K$-$A diagram.  The ISO/SWS spectra
\citep{Kraemer2002} are convolved with the MSX A-band filter transmission
curve.  The zero magnitude for MSX is taken from \citet{Cohen2000}.  The
ISO/SWS stars are cross-correlated with the 2MASS data base and SIMBAD for
identifications.  Selection criteria are $50\,{\rm Jy}<F({\rm A})<10000\,{\rm
  Jy}$ for ISO/SWS, and 2MASS flags either a,b,c,d in all of the JHK bands.
This restricts the Galactic sample to 271 spectra. Some stars were observed
several times and all of the A-band magnitudes were analysed individually.
Spectral types range from B-type to M- and C-types, and post-AGB stars and PNe
(all of them are WC), as well as young stellar objects are included.  The
M-type stars are all giants, sub-super giants, or super giants; no
dwarfs or sub-dwarfs are selected.

The J$-$K vs K$-$A diagram for this Galactic sample is shown in Fig.
\ref{iso.ps}.  Early-type stars have increasing J$-$K while K$-$A remains
constant, due to continuum H$^-$ absorption. This H$^-$ absorption is still
important for early M-type stars, while circumstellar excess affects both
colours among late M-type stars.  C-type refers to spectroscopically known
carbon stars; these form a relatively narrow sequence. The isolated carbon
star at [K$-$A]$\sim 15$ is AFGL 5625: it shows a feature-less spectrum (i.e.
no SiC), except for 14-$\mu$m C2H2. The OH/IR stars show a large range of
colours; OH/IR stars with [K$-$A]$\sim 4-10$ show silicate emission with
self-absorption on top (e.g., CRL 2199, WX Psc), whilst OH/IR stars with
[K$-$A]$>10$ show silicate absorption.

The separation between C-type and M-type stars with low mass-loss rates
($2<{\rm K-A}<3$) is due to photospheric, near-infrared molecular bands.  For
oxygen-rich Mira stars, the Ks-flux is suppressed by water absorption
\citep[e.g.,][]{Tej2003, Matsuura2002b}. For carbon stars, the J-band flux is
suppressed by C$_2$ and CN absorption \citep{Loidl2001}. At high mass-loss
rates, the carbon stars have well-defined colours which overlap with the (more
scattered) oxygen-rich stars.  Extremely red stars have no counterpart in the
LMC sample of \citet{Egan2001}.  This may be because of the 2MASS detection
limit.

At high mass-loss rates, the Egan et al. diagram cannot distinguish carbon and
oxygen-rich stars on an individual basis. However, stars on the carbon-star
sequence are statistically more likely to be carbon stars. Our selected stars
fall on this sequence. At low mass-loss rates the Egan et al. colours of our
targets are consistent with a carbon-rich nature. The sole oxygen-rich AGB
star, IRAS 05003$-$6712, has colours consistent with the carbon-rich sequence
(J$-$K$=2.95$, K$-$A$=3.8$), but at the blue end of the range where the two
types overlap.

\subsection{Molecules}
\begin{figure}
\includegraphics[width=8cm,clip=true]{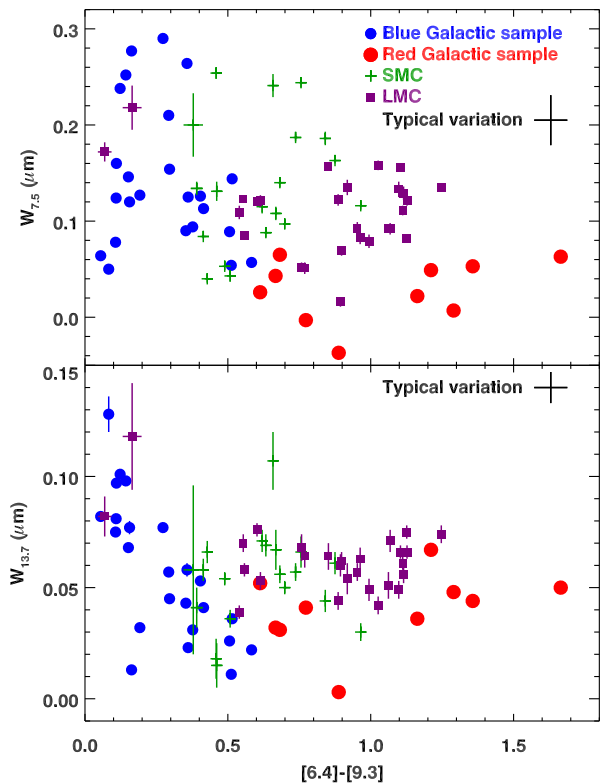}
\caption{\label{c2h2_col.eps} The strength of the C$_2$H$_2$ features
  (7.5$\mu$m and 13.7$\mu$m), as
     function of the [6.4]-[9.3] colour. Symbols are the same as in 
     Fig.~\ref{dustbands.ps}.}
\end{figure}

All molecules suggested to be present (C$_3$, HCN, C$_2$H$_2$) are so-called
parent molecules, which form in the photosphere, and are found throughout the
outflow, up to the photodissociation region at $\approx$ 1--3\,$ 10^{16}\,\rm
cm$ \citep{Millar1994, Millar2001}. Especially C$_2$H$_2$ is fundamental in
the chemistry leading to large organic molecules such as PAHs.  The abundance
of these molecules is an important parameter for our understanding of
chemistry in low-metallicity shells.  \citet{vanLoon1999b} noticed that the
3.1-$\mu$m band in LMC stars was of the same strength as in Galactic stars,
and suggested that the lower metallicity was compensated by a higher C/O
ratio. \citet{Matsuura2002a, Matsuura2005a} find that C$_2$H$_2$ is more
abundant in LMC and SMC stars than in Galactic comparison stars, based on the
3.1 and 3.8-$\mu$m absorption bands. Chemical models \citep{Matsuura2002a}
confirm the sensitivity of C$_2$H$_2$ to the C/O ratio. The carbon enhancement
from third dredge-up leads to a higher C/O ratio for lower initial oxygen
abundance.  \citet{vanLoon2006a} confirm that HCN is weak compared to
C$_2$H$_2$ and that CS is absent, using the shape and structure in the
3.8-$\mu$m absorption. They also show that the bands become sharper for stars
with stronger pulsation, and interpret this as evidence for cold molecular gas
at some distance above the stellar photosphere.

The equivalent widths of the molecular bands are listed in Table \ref{ew.dat}.
Fig.~\ref{c2h2_col.eps} plots these values for the C$_2$H$_2$ 7.5\,$\mu$m and
13.7\,$\mu$m bands as a function of the [6.4]$-$[9.3] colour. The squares in
the panels indicate the LMC stars. Especially the 13.7-$\mu$m band shows a
narrow range in equivalent width. The 7.5-$\mu$m band show a large range, but
neither correlates well with infrared colour.  The two 'naked' carbon stars
show stronger bands. The equivalent width of the 13.7-$\mu$m feature excludes
the wings of the broad component, and may therefore underestimate its strength
for hot (photospheric) gas.

\citet{Sloan2006} compare the strengths of these features for Galactic carbon
stars and SMC carbon stars.  They find that for stars with [6.4]$-$[9.3]$>$
0.6, the C$_2$H$_2$ bands are stronger in the SMC than in the Galaxy, the
difference being more readily apparent at 7.5\,$\mu$m than at 13.7\,$\mu$m.
The difference of strength between Galactic, LMC and SMC stars is interpreted
as due to higher abundances of C$_2$H$_2$ in the SMC, in accordance with the
prediction of \citet{Matsuura2005a}. Fig. \ref{c2h2_col.eps} includes the
Galactic and SMC samples. The equivalent widths are measured in the same way
for all stars. The Galactic sample (based on ISO spectra) is divided in 'blue'
and 'red' stars, based on whether the [6.4]$-$[9.3] colour is bluer than the
lowest extinction DUSTY model (Section \ref{cont_models}): blueward colours
are indicative of naked carbon stars.

The Galactic sample has a large range of [6.4]$-$[9.3] colour. The SMC sample
in contrast shows a very limited colour range.  For the same mass-loss rate
one may expect lower metallicity stars to have lower optical depths because of
a lower dust abundance. The SMC stars should show a lower [6.4]$-$[9.3] than
the LMC stars, whilst the Galactic stars should show the reddest colours.  This
effect is best seen  for [6.4]$-$[9.3]$>0.9$. In the models of
Section \ref{cont_models}, an increase of a factor of 3 in optical depth
reddens the [6.4]$-$[9.3] from 0.8 to 0.9, from 1.05 to 1.3, and from 1.3 to
1.8. The metallicity effect on colour is therefore much stronger for the
redder stars, consistent with the distribution in the figure.

The bluest objects, which are expected to be naked carbon stars, are almost
exclusively Galactic stars, with two LMC objects.  This is a selection effect,
in that only mass-losing stars were selected for the various Magellanic Cloud
surveys. For the redder stars, the shift in colour with optical depth suggests
the Galactic stars with [6.4]$-$[9.3] between 1.2 and 1.8 should be compared
with the LMC stars between 1.0 and 1.2, and the SMC stars around 0.9.  In the
top panel, this shows a significant relation, with the 7.5-$\mu$m feature
becoming much stronger for lower metallicity. For the bluer mass-losing stars,
where there is little change of colour with optical depth, the LMC stars have
a stronger 7.5-$\mu$m band than Galactic stars, but there is no clear offset
between LMC and SMC stars.

Overall, this provides some evidence for the suggestion that the C$_2$H$_2$
bands are stronger in lower metallicity environments. For the 13.5-$\mu$m
band, a small shift is also seen between the Galactic and the LMC stars,
although less convincing than at 7.5$\mu$m, but the LMC and SMC stars show
comparable equivalent widths.

The fact that the C$_2$H$_2$ bands show a relatively small range in equivalent
width for a large range of colour temperatures, confirms that the molecule is
located throughout the shell.

\subsection{Dust}

\subsubsection{Temperature dependence}
\label{dustband_section}

\begin{figure*}
\includegraphics[width=16cm,clip=true]{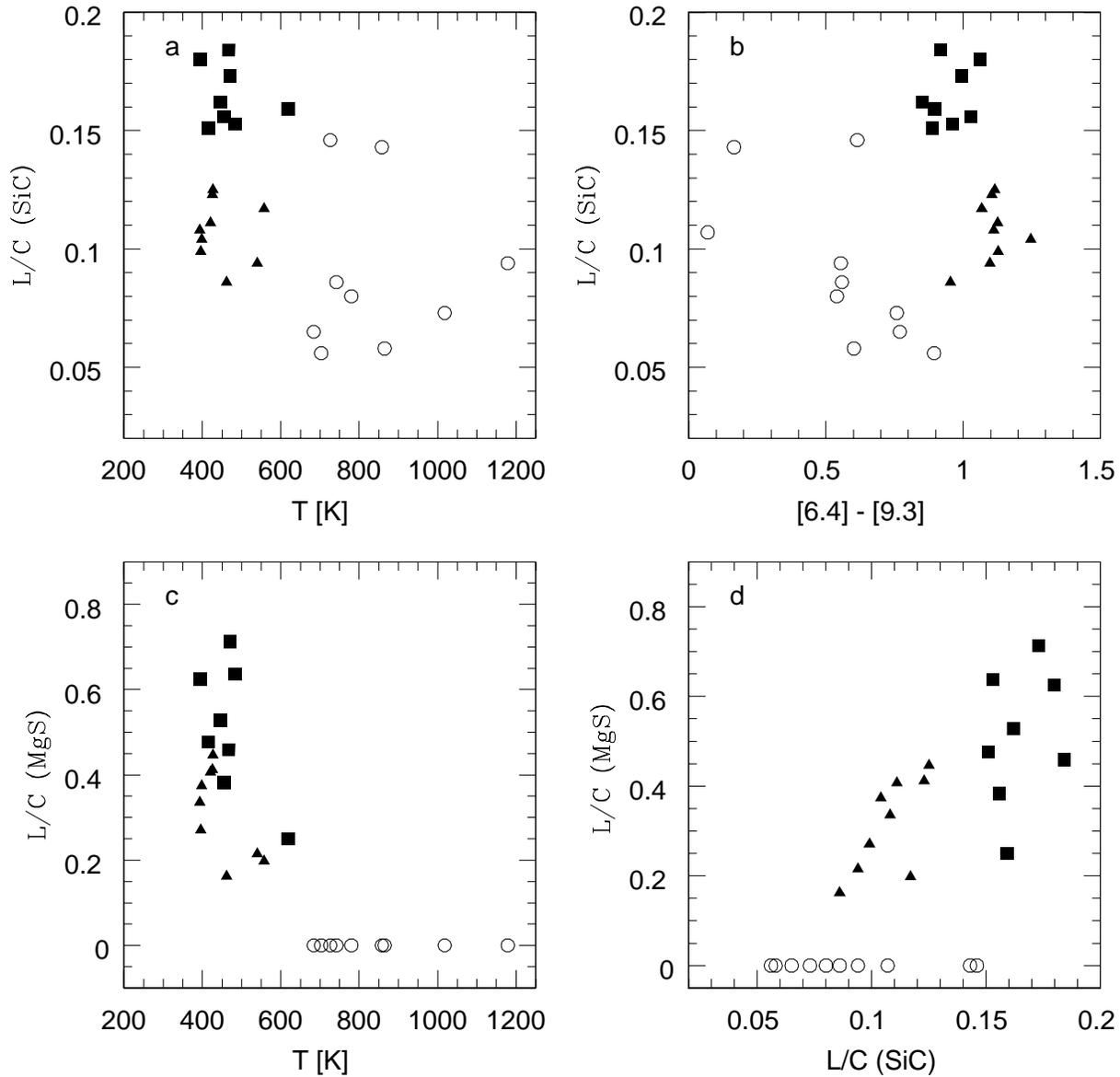}
\caption{\label{dustbands.ps} The strength of the 11-micron SiC and the
  30-micron MgS bands, as function of the dust continuum temperature (derived
from [16.5]$-$[21.5]), and the [6.4]$-$[9.3] colour. Circles indicate
stars without an MgS feature; Squares: stars with an MgS feature and 
a strong SiC feature; Triangles: stars with an MgS feature and
a weak SiC feature.}
\end{figure*}

The {\it Spitzer} spectra show that SiC dust bands are observed for most or
all carbon stars, but the MgS band is present in only a subset. This already
indicates that the MgS mineral forms under more restricted conditions. The
occurrence and strength of the two bands is plotted for the LMC stars in Fig.
\ref{dustbands.ps}, as function of the continuum temperature (derived from
[16.5]$-$[21.5]) and of the [6.4]$-$[9.3] colour, as tracer of optical depth.
The sources are put in three different groups: the circles indicate stars
without an MgS feature; triangles show stars with MgS and with weak SiC
(defined as a line-to-continuum ratio less than 0.14); squares show stars with
MgS and stronger SiC.

The figure shows that the presence of MgS correlates strongly with the dust
temperature. There is a sharp division at a (black body) temperature of
$T=650\,$K: none of the stars with hotter dust show MgS, whilst all stars with
cooler dust do. Between dust temperatures of 500 and 600\,K the MgS feature is
relatively weak, and at lower temperatures the feature is stronger and shows a
large range of line-to-continuum ratio. In contrast, SiC tends to be weaker at
high dust temperatures but is almost always present.  Panel d indicates that
increasing SiC strength is accompanied by rapidly increasing MgS. The
difference can be understood in terms of the formation mechanism of the
different dust species.  Solid SiC condenses directly from the gas, at high
temperatures. MgS grows as a surface on pre-existing grains \citep{Nuth1985}:
this process only starts at temperatures around 600\,K, and is complete around
300\,K.

\begin{figure}
\includegraphics[width=7cm]{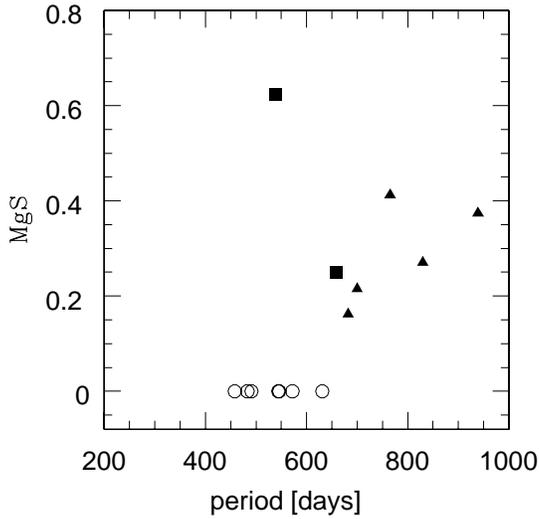}
\caption{\label{periods.ps} The MgS line-to-continuum ratio as function of
pulsation period of the star. Open circles: no MgS, triangles: weak SiC;
squares: strong SiC (Fig \ref{dustbands.ps}) }
\end{figure}

Fig. \ref{periods.ps} shows a less common way to look at MgS, using the
relation with pulsation period of the star (taken from Table
\ref{periods.dat}). The double-period star is not plotted. The MgS feature is
seen in general for stars with periods longer than 650 days. The relation is
an indirect one, in the sense that the longest period stars tend to have the
highest mass-loss rates.

\subsubsection{Metallicity dependence}

\begin{figure}
\includegraphics[width=8cm,clip=true]{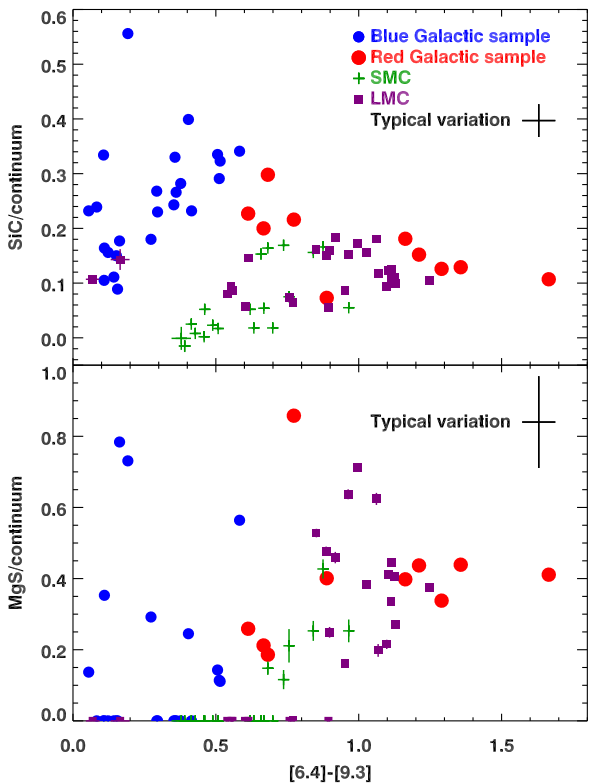}
\caption{\label{dust_strength} The strength of the SiC and MgS features as a
     function of the [6.4]-[9.3] colour.}
\end{figure}

Fig.~\ref{dust_strength} compares the strengths of the SiC and MgS features in
our LMC sample with the SMC and Galactic samples of \citet{Sloan2006}. Sloan
et al.  find that the SiC and MgS features are weaker in the SMC than in the
Galaxy. This is expected, as Mg, Si and S are not produced in AGB stars and
their abundance should scale directly with metallicity. Our data show a more
complicated picture, however.

The line-to-continuum ratios are plotted against [6.4]$-$[9.3], which is a
measure of optical depth. As before, for the higher-mass-loss stars the
optical depth is less in the SMC, while at low optical depth SMC, LMC and
Galactic stars can be compared at the same value for [6.4]$-$[9.3].  The top
panel shows that at low optical depth ([6.4]$-$[9.3]\,$\approx 0.5$), there is
a clear sequence with the SiC line-to-continuum ratio diminishing with
decreasing metallicity. At higher optical depth, the three samples indicate
very similar ratios. Stars with 'naked carbon star' colours show a high
line-to-continuum ratio, related to the weakness of the continuum, with a
sequence clearly offset from the dusty stars. It is possible that here the
continuum used to measure the band strength is affected by molecular
absorption bands, giving an artifically enhanced feature. The SiC band is
flanked by two photospheric bands, attributed to C$_2$H$_2$ and C$_3$, and a
true continuum cannot easily be defined.

The MgS feature is illustrated in the bottom panel of Fig.
\ref{dust_strength}.  There is an indication that the MgS band is weaker in
the SMC than in comparable Galactic and LMC stars. The LMC stars show a very
large scatter and it is not possible to draw conclusions regarding its typical
strength. We note that the continuum underneath the feature is not well
defined in {\it Spitzer} spectra because of its long-wavelength cutoff.

In both cases the line-to-continuum ratio in dusty stars compares the band
strength with the underlying dust continuum, assumed to be due to amorphous
carbon. As the carbon is dredged-up in the star while the mineral constituents
are not, the ratio is expected to decrease with decreasing metallicity.
However, this is also affected by the dust condensation process. If seeds are
needed to grow the amorphous carbon grains (e.g. TiC), the relation between
mineral and amorphous carbon dust could be more complicated. Especially the
SMC data indicate that the emissivity ratio between minerals and amorphous
carbon dust is lower at low metallicity, but the LMC data is inconclusive on
this point.

The abundance of SiC is limited by the enriched carbon and the unenriched
silicon. The abundance of MgS is limited by two unenriched elements. If the
condensation is 100 per cent efficient, both will be affected similarly by
metallicity. If the condensation operates at lower efficiency, the formation
becomes limited by the collision rate and the the abundance of MgS would more
strongly dependent on metallicity than SiC ([MgS]$\propto Z^2$, while
[SiC]$\propto Z$, where $Z$ is the fractional abundance of metals).
(Temperature effects may also contribute, as lower opacity gives rise to
higher dust temperatures.) The spectra presented here do not support such a
strong dependence on metallicity, but favour a linear relation. This supports
a high condensation efficiency for MgS.  Confirmation will require detailed
modelling of the dust spectra.

\begin{figure}
\includegraphics[width=8cm,clip=true]{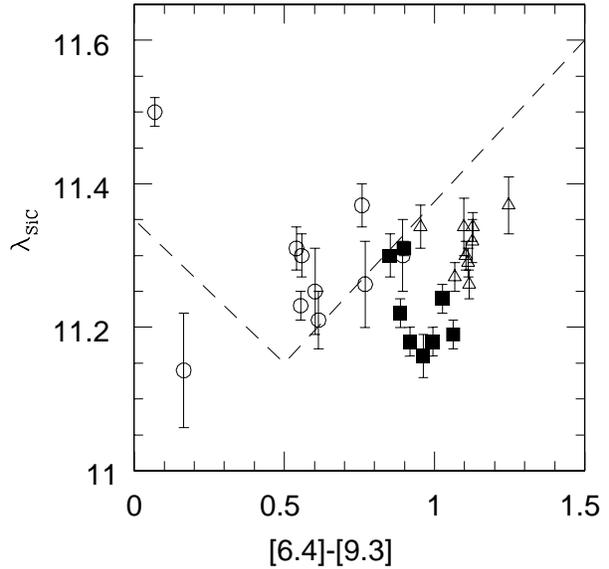}
\caption{\label{temp_lambda_sic_alb.ps} Apparent central wavelength of the SiC
  feature as a function of the [6.4]-[9.3] colour. The dashed line represents
  the distributions of Galactic carbon stars, according to \citet{Sloan2006}.
  Symbols are as in Fig.~\ref{dustbands.ps}.}
\end{figure}

\subsubsection{Central wavelengths}

The central wavelengths of the dust bands are important diagnostics
\citep{Speck2005, Hony2002b} of the dust environment. We define the central
wavelength (hereinafter $\lambda_{\rm SiC}$) as the wavelength which disects
the flux distribution in the band: i.e. after continuum subtraction, half the
band flux arises redward and half blueward. This definition is more stable
against noise on the spectral energy distribution than the wavelength of peak
emission strength. However, for MgS only part of the band is covered by the
{\it Spitzer} spectra and the derived central wavelength reflects this.

Fig.~\ref{temp_lambda_sic_alb.ps} shows the observed variation of
$\lambda_{\rm SiC}$, as a function of the [6.4]$-$[9.3] colour.
\citet{Sloan2006} have shown that for Galactic carbon stars with
[6.4]$-$[9.3]\,$>$0.5, $\lambda_{\rm SiC}$ increases monotonically with
[6.4]$-$[9.3] colour, from $\sim 11.20\,\mu$m up to $\sim 11.70\,\mu$m for the
reddest source of their sample.  For bluer colours $\lambda_{\rm SiC}$ reddens
again: their relation is indicated by the dashed line. \citet{Speck2005} argue
that red carbon-rich stars can show the SiC in self-absorption and
full-absorption.  This absorption occurs at 10.8 $\mu$m, so that an increase
of the amount of SiC dust leading to self-absorption will cause a redward
shift of $\lambda_{\rm SiC}$. The presence of an absorption component in all
stars remains to be proven.

The inverse shift to longer wavelength of $\lambda_{\rm SiC}$, for stars with
[6.4]$-$[9.3]\,$<$0.5, is not yet explained. This colour corresponds to naked
carbon stars, where the [9.3] band may be suppressed by a photospheric
molecular band. This band borders the SiC feature, and may overlap with it in
wavelength. This would have a similar effect as the self-absorption proposed
by \citet{Speck2005}, and could self consistently explain why a bluer
[6.4]$-$[9.3] colour (deeper 10-$\mu$m absorption) correlates with a redder
SiC feature.  Note that the broad molecular absorptions at 10 and 13\,$\mu$m
may give the impression of an intermediate emission band: the actual presence
of SiC in such stars may be in doubt.

In our LMC sample, we observe that $\lambda_{\rm SiC}$ is redder for the stars
with weaker SiC features (triangles in Fig.~\ref{temp_lambda_sic_alb.ps}).
These stars have slightly redder [6.4]$-$[9.3] as shown in Fig.
\ref{dustbands.ps}.  This can in principle be interpreted as being due to
self-absorption weakening the band, although other explanations are possible.
In our sample, the dust continuum temperature has no discernable influence on
$\lambda_{\rm SiC}$. 

The LMC stars with MgS (triangles and squares: cool dust) follow the slope of
the relation of \citet{Sloan2006}, but offset towards redder [6.4]$-$[9.3]. An
alternative view would be that at the same optical depth, LMC stars have a
bluer $\lambda_{\rm SiC}$. This may reflect the broad C$_2$H$_2$ R-band at 
12--14\,$\mu$m, depressing the apparent red continuum. A stronger  C$_2$H$_2$
band in LMC stars would shift the observed  $\lambda_{\rm SiC}$ to shorter
wavelength.

\begin{figure}
\begin{center}
\includegraphics[width=9cm,clip=true]{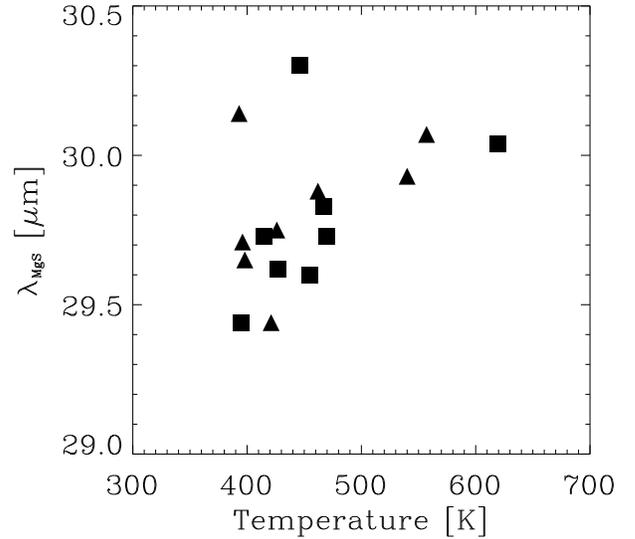}
\caption{\label{col_lambmgs_alb.ps}Apparent
  central wavelength of the MgS feature as a function of the temperature
  (derived from the [16.5]$-$[21.5] colour).  Symbols are  as in 
  Fig.~\ref{dustbands.ps}.}
\end{center}
\end{figure}

\begin{figure*}
\begin{center}
\includegraphics[width=15cm,clip=true]{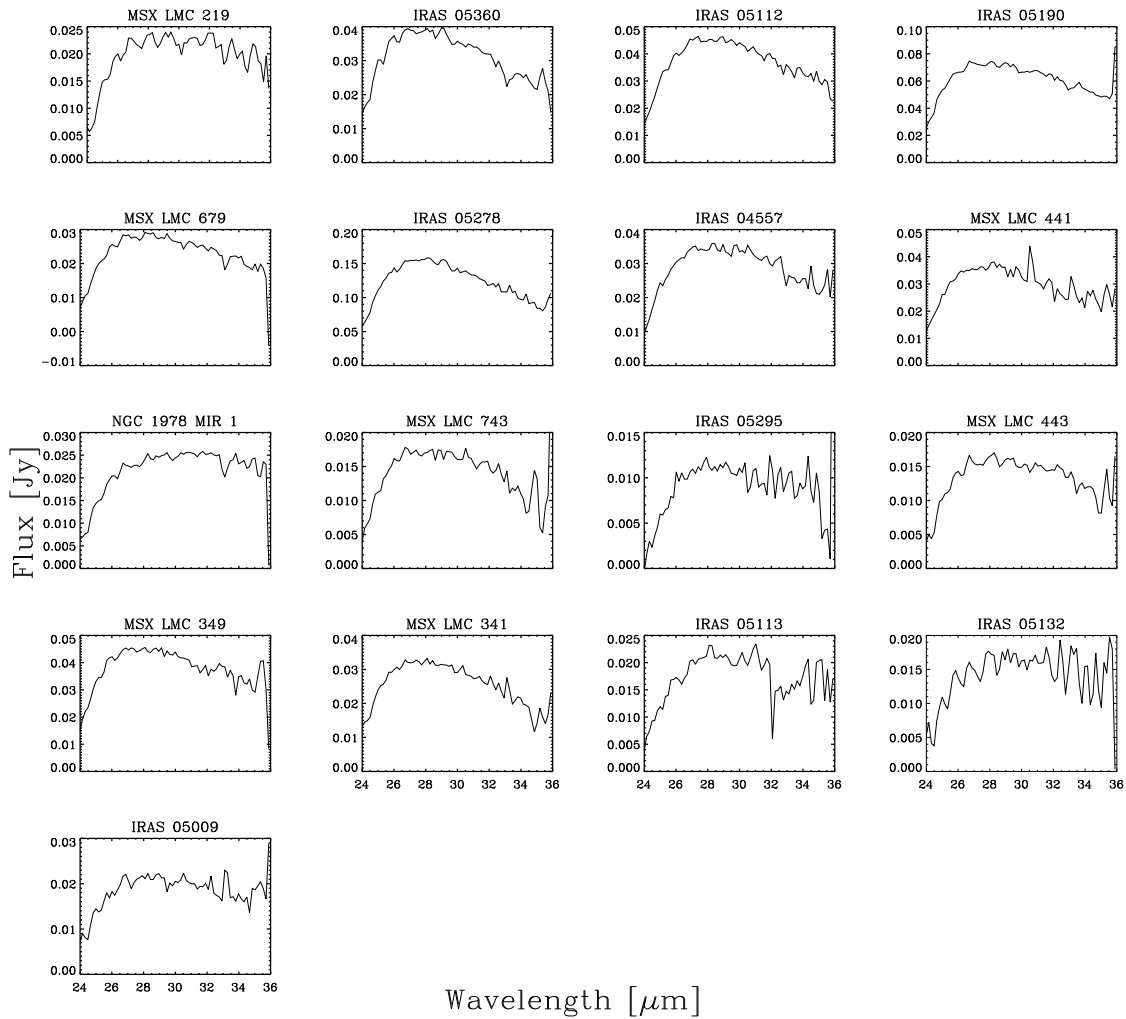}
\caption{\label{mgs_feature_temp.ps} Continuum subtracted shape of the MgS
  features observed in our LMC sample. The spectra are ordered by dust
  temperature, from blue spectra to red.}
\end{center}
\end{figure*}

Fig.~\ref{col_lambmgs_alb.ps} shows the central wavelength of the MgS feature
(hereinafter $\lambda_{\rm MgS}$, defined as for the SiC feature central
wavelength) as a function of the temperature derived from the [16.5]$-$[21.5]
colour. From a large sample of Galactic evolved stars, \citet{Hony2002b} show
that the central wavelength of the MgS feature is independent of continuum
temperatures. They explain this by the fact that MgS is partially heated by
mid-IR radiation. In our sample, the MgS band is observed only in stars spread
over a small range of continuum temperatures, making it difficult to find any
correlation between $\lambda_{\rm MgS}$ and the temperature. The suggestion of
longer $\lambda_{\rm MgS}$ for higher dust temperatures is not conclusive.
Our LMC sample shows $\lambda_{\rm MgS}$ similar to Galactic stars observed by
Hony et al.  within the same temperature range.
Fig.~\ref{mgs_feature_temp.ps} show the shape of the continuum-substracted MgS
feature. This figure confirms that there is no apparent correlation between
the MgS feature shape and the dust continuum temperature. The short-wavelength
band edge has a similar shape in all spectra. There is some separation between
stars with relatively flat MgS band and stars whhere the band diminishes
towards the red, but this separation does not show an evident correlation with
dust temperature.


\subsubsection{Oxygen-rich dust}

The current sample contains almost exclusively carbon-rich stars.  Oxygen-rich
mass-losing stars do exist in the LMC, as found in the sample of, e.g.,
\citet{Trams1999} \citep[see also ][]{Dijkstra2005}. Comparing the
J$-$K versus K$-$A diagram for MSX sources in \citet{Egan2001}, with
the diagrams in Section \ref{carbonclass}, shows that the majority of
cocoon LMC stars have colours consistent with the carbon star
sequence, but some stars are bluer in K$-$A and these could be oxygen-rich.

The higher efficiency of
third dredge-up at low metallicity turns LIMS into carbon stars relatively
early on the AGB: the existence of optical (i.e. unobscured) carbon stars show
that in the LMC this typically occurs before the onset of the high mass-loss
phase.  Oxygen-rich mass-losing stars will most likely be either stars with
initial masses large than about 4\,M$_\odot$, where hot bottom burning
prevents the star from becoming carbon-rich, or low-mass stars.

Dust formation in oxygen-rich stars is dependent on metallicity-dependent
minerals, such as corundum and various silicates.  Oxygen-rich stars may
therefore show relatively poor dust formation efficiency at low metallicity.
For some stars, this could in principle delay the superwind phase sufficiently
to allow them to become a carbon star. Carbon stars are less affected by
metallicity, as dust is produced from self-enriched carbon.  It would be of
interest to test whether the LMC stars show any evidence for a delayed or
weaker superwind for oxygen-rich stars.

The predominance of mass-losing carbon stars implies that dust enrichment of
the ISM at low metallicity by AGB stars will be strongly dominated by
amorphous carbon dust, as compared to the Galaxy.

\subsection{Three groups}

In Sect.~\ref{dustband_section}, we have shown that the plot of the strength
of the dust bands versus colours or temperatures permits us to identify three
distincts group of carbon stars in our LMC sample. The three groups can be
defined as: (1) stars with a strong SiC feature and MgS, (2) stars with a weak
SiC feature and MgS, and (3) stars without an MgS feature.

\begin{figure*}
\includegraphics[width=16cm,clip=true]{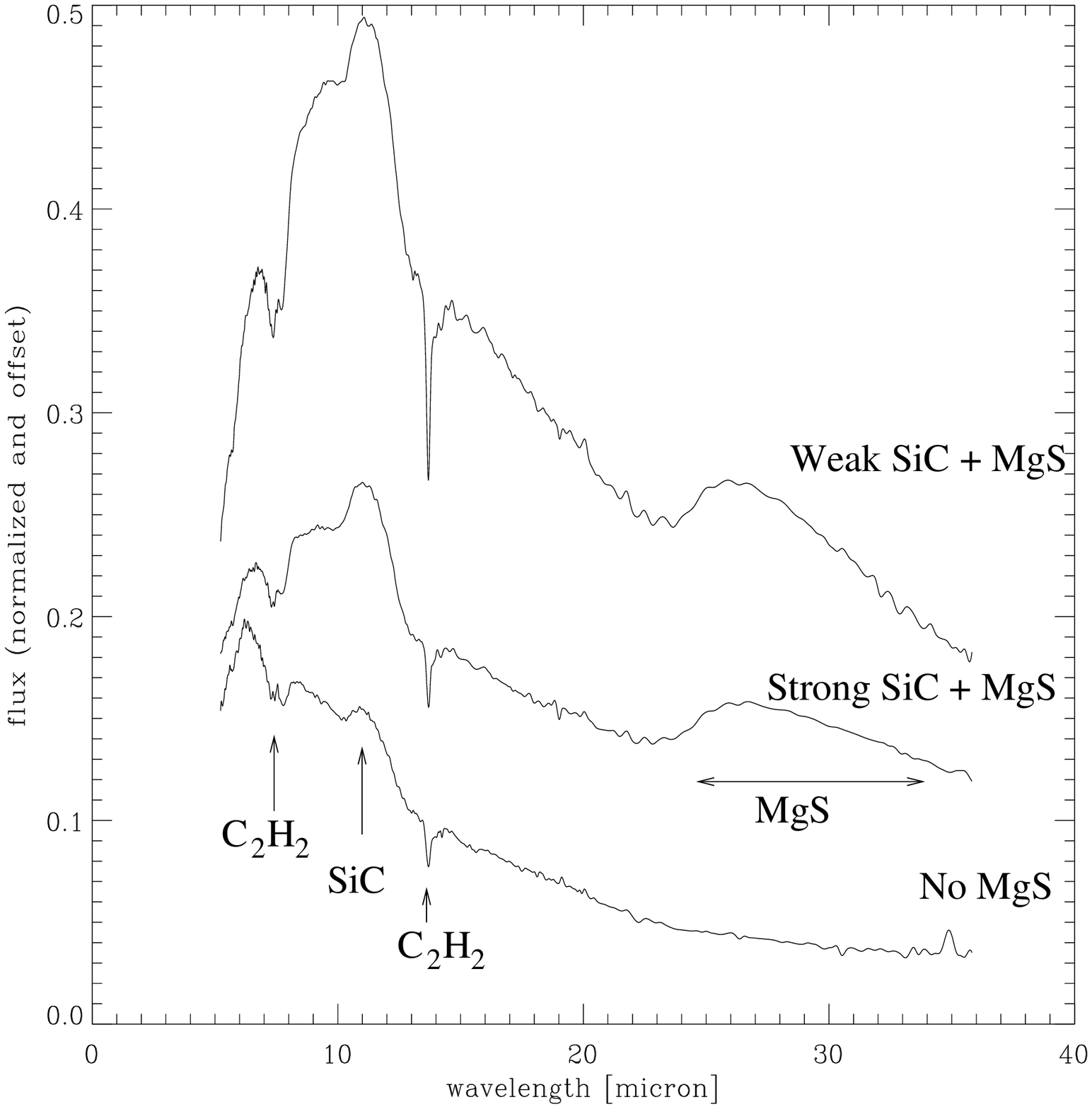}
\caption{\label{average_spec.eps} Spectra of the three object classes
  observed, the first class (top) containing objects displaying a weak SiC
  feature and a MgS feature, the second (middle) a strong SiC and a MgS
  feature, the last one (bottom) displaying no MgS feature. These three
  spectra are obtained by averaging the individual spectra for each class. }
\end{figure*}

Fig.~\ref{average_spec.eps} shows the spectral shapes of stars within those
three groups, obtained by averaging the different spectra within each group.

The shape of the SiC band is essentially identical between the first two
groups. The continuum on either side is different. The narrow 13.7-$\mu$m band
weakens as SiC becomes weaker, but the broad band between 12 and 16\,$\mu$m
becomes much more prominent. This is best seen by comparing the slopes of the
spectra. The blue slopes of the continua (5--9\,$\mu$m) also show a difference
between the three groups.

Such a classifiation can be compared with that proposed by
\citet{Kraemer2002}, who divide dusty carbon-rich stars into red (CR) and blue
(CE). In our sample, about 10 stars would be classified as CE and the
remainder (apart from the symbiotic C star) as CR. In the Galactic ISO sample
of \citet{Kraemer2002}, CE stars are somewhat more dominant. This difference
is likely to be caused by selection effects.

\section{Conclusions}

We have presented a {\it Spitzer} survey of 29 mass-losing Asymptotic Giant
Branch stars in the Large Magellanic Cloud.  Most are classified as carbon
stars, based on the infrared spectra. One of the C stars is also a symbiotic
star. Two oxygen-rich stars were also present in the original survey: one was
found to be a high-mass young stellar object and is presented in
\citet{vanLoon2005b}. 

The spectra cover the wavelength range 5--38\,$\mu$m. The short wavelength
range, up to 16\,$\mu$m, is dominated by absorption bands caused by
circumstellar and/or photospheric molecules. At the short edge, bands of CO
and C$_3$ cause a sharp drop of the stellar energy distribution towards the
blue end of the spectral range. Absorption bands at 7.5 and 13.7\,$\mu$m are
due to acetylene, C$_2$H$_2$. The 13.7$\mu$m band contains both a narrow and a
broad component, the latter tracing hotter gas. These bands are also seen in
two naked carbon stars in our sample.

Dust emission bands are present in our sample at 11 and at 30\,$\mu$m, due
respectively to SiC and MgS. The first band is present in almost every object,
regardless of dust temperature. MgS is only present when the dust temperature
is below 650\,K, approximately in half the sample. This is consistent with how
these minerals form: SiC condenses directly out of the gas phase, but MgS is
deposited on existing dust grains below a critical temperature.

Because of the many bands in the spectrum, continuum definition is difficult.
We define a set of four narrow bands, called the Manchester system, which can
be used to trace the continuum for dusty carbon stars. The bands are centred
at 6.4, 9.3, 16.5 and 21.5\,$\mu$m.  The [6.4]$-$[9.3] colour is shown to be a
measure of the optical depth in the shell. The [16.5]$-$[21.5] colour can be
used to measure the dust temperature.  Two naked carbon stars show much bluer
colours in [6.4]$-$[9.3] than expected from models. This is attributed to
suppression of the [9.3] band by a molecular band, which is apparent in many
of the stars but is particularly clear in stars with photospheric infrared
emission. A 10-$\mu$m absorption band in carbon stars has previously been
attributed to interstellar silicate, but this assignment can be ruled our for
the LMC stars. A recent assignment to silicon-nitrite particles can also be
ruled out for emission at photospheric temperatures. This supports the
identification with a C$_3$ band, as proposed by \citet{Jorgensen2000}.

The classification scheme of mass-losing stars by \citet{Egan2001} is
investigated. Based on this scheme, some stars in our sample were expected to
be oxygen-rich. We show from current data, from dust models and from ISO
spectra of Galactic stars, that for high mass-loss stars, the J$-$K, K$-$A
colours cannot easily distinguish oxygen-rich from carbon-rich stars.  The
fact that almost all our sample was found to be carbon rich reflects the fact
that they lie on a sequence of increasing mass-loss rate, with the stars
becoming carbon stars even before they develop thin dust shells.

We investigate the strengths of the molecular and dust bands as function of
metallicity, based on a comparison with carbon stars in the Galaxy and the
SMC. The 7.5-$\mu$m C$_2$H$_2$ band shows evidence for increasing equivalent
width with decreasing metallicity. The 13.7-$\mu$m band does not clearly show
this.  The SiC band shows evidence for a much lower line-to-continuum ration
at the lowest metalicity (the SMC) but no strong difference between the LMC
and the Galaxy. For MgS there is an indication for a sequence of decreasing
line-to-continuum ratio with decreasing metallicity. We show that the central
wavelength of the SiC feature depends on the [6.4]$-$[9.3] colour, consistent
with \citet{Sloan2006}. It is however important to realize the uncertainty of
the continuum determination, affected by the molecular absorption bands on
either side.

The increasing strength of acetylene bands with decreasing metallicity can be
explained by the effect of carbon dredge-up: a fixed amount of added carbon
gives a higher C/O ratio at lower metallicity (less O), leaving more free
carbon after the formation of CO. This may also be the cause of the C$_3$
bands which we see. SiC and MgS do not benefit form this effect, and their
abundance is expected to scale more directly with metallicity.

We finally note that dust formation in oxygen-rich stars is dependent on
metallicity-dependent minerals, such as corundum and various silicates.
Oxygen-rich stars may therefore show relatively poor dust formation efficiency
at low metallicity. Carbon stars, on the other hand, depend for dust on
self-produced carbon and can have high efficiency of forming amorphous carbon
dust.  The earlier formation of carbon stars due to third dredge-up, together
with a possible difference in dust formation efficiency, means that dust
enrichment of the ISM at low metallicity by AGB stars will be dominated by
amorphous carbon dust, as compared to the Galaxy.

\section*{Acknowledgements}

We appreciate the efforts of the IRS team at Cornell. AAZ, EL and MM
acknowledge PPARC rolling grant support. PPARC also supported this research
via a visitor grant. PRW acknowledges the support provided by a grant from the
Australian Research Council. MM thanks the UMIST Peter Allen Travelling Grant.
M. Cohen kindly provided us with the MSX filter transmission curve.

\bibliographystyle{mn2e}

\bibliography{paper-overview}

\end{document}